\documentclass[11pt,a4paper]{emulateapj}
\usepackage{graphicx}
\usepackage{textcomp}
\usepackage{amsmath}
\usepackage{amssymb}
\usepackage{gensymb}
\usepackage{rotating}
\usepackage{epsfig}
\usepackage{amsmath}
\usepackage{longtable}
\usepackage{color}
 \def\apjl{ApJL}

 \def\aap{A\&A}
 
 \def\apjs{ApJS}

 \def\gs{\mathrel{\raise0.35ex\hbox{$\scriptstyle >$}\kern-0.6em\lower0.40ex\hbox{{$\scriptstyle \sim$}}}}
 \def\ls{\mathrel{\raise0.35ex\hbox{$\scriptstyle <$}\kern-0.6em\lower0.40ex\hbox{{$\scriptstyle \sim$}}}}

 \def\Msol{\mathrel{\rm M_{\odot}}}
 \def\Lsol{\mathrel{\rm L_{\odot}}}
 
 \def\Wm2{\,\hbox{W}\,\hbox{m}^{-2}}
 \def\gsim{\mathrel{\raise0.35ex\hbox{$\scriptstyle >$}\kern-0.6em\lower0.40ex\hbox{{$\scriptstyle \sim$}}}}
 \def\lsim{\mathrel{\raise0.35ex\hbox{$\scriptstyle <$}\kern-0.6em\lower0.40ex\hbox{{$\scriptstyle \sim$}}}}
 
 \def\pc{\%}

\lefthead{Simpson et al.}  \righthead{S2CLS: Resolved ALMA imaging of sub--millimeter galaxies}

\begin{document}

\title{The SCUBA-2 Cosmology Legacy Survey: ALMA resolves the
  rest--frame far-infrared emission of sub-millimeter galaxies}

\author{
J.\,M.\ Simpson,\altaffilmark{1}
Ian Smail,\altaffilmark{1}
A.\,M.\ Swinbank,\altaffilmark{1}
O.\ Almaini,\altaffilmark{2}
A.\,W.\ Blain,\altaffilmark{3}
M.\,N.\ Bremer,\altaffilmark{4}
S.\,C.\ Chapman,\altaffilmark{5}
Chian-Chou\ Chen,\altaffilmark{1}
C.\ Conselice,\altaffilmark{2}
K.\,E.\,K.\ Coppin,\altaffilmark{6}
A.\,L.\,R.\ Danielson,\altaffilmark{1}
J.\,S.\ Dunlop,\altaffilmark{7}
A.\,C.\ Edge,\altaffilmark{1}
D.\ Farrah,\altaffilmark{8}
J.\,E.\ Geach,\altaffilmark{6}
W.\,G.\ Hartley,\altaffilmark{2,9}
R.\,J.\ Ivison,\altaffilmark{7,10}
A.\,Karim,\altaffilmark{11}
C.\,Lani,\altaffilmark{2}
C.\,--J.\ Ma,\altaffilmark{1}
R.\,Meijerink,\altaffilmark{12}
M.\,J.\ Micha{\l}owski,\altaffilmark{7}
A.\ Mortlock,\altaffilmark{2,7}
D.\ Scott,\altaffilmark{13}
C.\,J.\ Simpson,\altaffilmark{14}
M.\ Spaans,\altaffilmark{15}
A.\,P.\ Thomson,\altaffilmark{1}
E.\ van Kampen,\altaffilmark{10}
P.\,P.\ van der Werf \altaffilmark{12}
}

\setcounter{footnote}{0}
\altaffiltext{1}{Institute for Computational Cosmology, Department of Physics, Durham University, South Road, Durham DH1 3LE, UK; email:   j.m.simpson@dur.ac.uk}
\altaffiltext{2}{School of Physics and Astronomy, University of Nottingham, Nottingham NG7 2RD, UK}
\altaffiltext{3}{Department of Physics \& Astronomy, University of Leicester, University Road, Leicester LE1 7RH, UK}
\altaffiltext{4}{School of Physics, HH Wills Physics Laboratory, Tyndall Avenue, Bristol BS8 1TL, UK}
\altaffiltext{5}{Department of Physics and Atmospheric Science, Dalhousie University Halifax, NS B3H 3J5, Canada}
\altaffiltext{6}{Centre for Astrophysics Research, Science and Technology Research Institute, University of Hertfordshire, Hatfield AL10 9AB, UK} 
\altaffiltext{7}{Institute for Astronomy, University of Edinburgh, Royal Observatory, Blackford HIll, Edinburgh EH9 3HJ, UK}
\altaffiltext{8}{Department of Physics, Virginia Tech, Blacksburg, VA 24061, USA }
\altaffiltext{9}{ETH Z{\"u}rich, Institut f{\"u}r Astronomie, HIT J 11.3, Wolfgang-Pauli-Str. 27, CH-8093 Z{\"u}rich, Switzerland }
\altaffiltext{10}{European Southern Observatory, Karl Schwarzschild Strasse 2, Garching, Germany}
\altaffiltext{11}{Argelander-Institute for Astronomy, Bonn University, Auf dem H{\"u}gel 71, D-53121 Bonn, Germany}
\altaffiltext{12}{Leiden Observatory, Leiden University, P.O. Box 9513, NL-2300 RA Leiden, Netherlands}
\altaffiltext{13}{Department of Physics \& Astronomy, University of British Columbia, 6224 Agricultural Road, Vancouver, BC, V6T 1Z1, Canada}
\altaffiltext{14}{Astrophysics Research Institute, Liverpool John Moores University, Liverpool Science Park, 146 Brownlow Hill, Liverpool L3 5RF, UK}
\altaffiltext{15}{Kapteyn Astronomical Institute, University of Groningen, The Netherlands}

\begin{abstract} 
We present high-resolution (0.3$''$) ALMA 870\,$\mu$m imaging of 52 sub-millimeter galaxies (SMGs) in the Ultra Deep Survey (UDS) field and investigate the size and morphology of the sub--millimeter (sub--mm) emission on 2--10\,kpc scales. We derive a median intrinsic angular size of FWHM\,=\,$0.30 \pm 0.04''$ for the 23 SMGs in the sample detected at a signal-to-noise ratio (SNR) $>10$. Using the photometric redshifts of the SMGs we show that this corresponds to a median physical half--light diameter of $2.4 \pm 0.2$\,kpc. A stacking analysis of the SMGs detected at an SNR\,$<10$ shows they have sizes consistent with the 870\,$\mu$m--bright SMGs in the sample. We compare our results to the sizes of SMGs derived from other multi--wavelength studies, and show that the rest--frame $\sim$\,250\,$\mu$m sizes of SMGs are consistent with studies of resolved $^{12}$CO ($J$\,=\,3--2 to 7--6) emission lines, but that sizes derived from 1.4\,GHz imaging appear to be approximately two times larger on average, which we attribute to cosmic ray diffusion. The rest--frame optical sizes of SMGs are around four times larger than the sub-millimeter sizes, indicating that the star formation in these galaxies is compact relative to the pre-existing stellar distribution. The size of the starburst region in SMGs is consistent with the majority of the star formation occurring in a central region, a few kpc in extent, with a median star formation rate surface density of 90\,$\pm$\,30\,$\Msol$\,yr$^{-1}$\,kpc$^{-2}$, which may suggest that we are witnessing an intense period of bulge growth in these galaxies . 
\end{abstract}

\keywords{galaxies: starburst, galaxies: high-redshift}

\section{Introduction}\label{sec:intro}

Nearly twenty years after their discovery, there is still debate about the nature of the population of luminous, but highly dust-obscured sources detected at high redshifts in sub-millimeter and millimeter surveys. The observational data suggest that the 850\,$\mu$m--detected  sub-millimeter galaxies (SMGs) lie at a median redshift $z$\,=\,$2.5$\,$\pm$0.2 (\,\citealt{Simpson14}, see also~\citealt{Chapman05,Wardlow11,Yun12,Smolcic12,Weiss13}) and are powered by bursts of star formation in relatively massive, gas-rich galaxies (stellar masses of $\sim$\,10$^{11}$\,M$_\odot$ and gas masses of $\sim$\,0.5\,$\times$\,10$^{11}$\,M$_\odot$, e.g.\ \citealt{Hainline11,Michalowski12,Bothwell13}) with space densities of $\sim$\,10$^{-5}$\,Mpc$^{-3}$. A modest proportion of SMGs have been shown to host an accreting super-massive black hole (e.g.\ \citealt{Alexander08,Pope08,WangS13}) and many appear disturbed or irregular in high-resolution rest--frame optical imaging from {\it HST}, albeit predominantly with a low S\'{e}rsic index (e.g. \citealt{Conselice03b,Chapman03b,Swinbank10,Targett13,Wiklind14,Chen14}). SMGs thus share some of the traits of local Ultraluminous Infrared Galaxies (ULIRGs), although they are $\sim$\,10$^3$ times more abundant at a fixed far-infrared luminosity (e.g.\ \citealt{Chapman05,Lindner11,Magnelli12,Yun12,Swinbank13}) and appear to be more massive than these proposed analogs (e.g.\,\citealt{Tacconi02}).

In the past decade, near--infrared (NIR) spectroscopy has also identified a population of quiescent, red, galaxies at $z$\,=\,1.5--3, which have been proposed as the potential descendants of high redshift starbursts (SMGs). The stellar populations in these high redshift quiescent galaxies follow a fairly tight ``red-sequence'', indicating that the stellar population was formed rapidly in an intense starburst phase (e.g.\ \citealt{Kriek08b}). The high star formation rates of SMGs (300\,$\Msol$\,yr$^{-1}$; \citealt{Magnelli12,Swinbank13}), combined with large molecular gas reservoirs, indicate that they have the potential to form a stellar component of $10^{10}$--10$^{11}$\,$\Msol$ in 100\,Myrs. Such rapid stellar mass growth, at high redshift, has led to speculation that SMGs may be the progenitors of both these high-redshift quiescent galaxies, and local elliptical galaxies~\citep{Lilly99, Genzel03, Blain04a, Swinbank06b, Tacconi08, Hainline11, Hickox12, Toft14, Simpson14}.

Studies investigating this proposed evolutionary scenario typically compare properties such as the stellar mass, spatial clustering and space densities of the population and the proposed descendants. However, each of these methods has significant associated uncertainties. In particular, the stellar masses of SMGs have been shown to be highly dependent on the assumed star formation history, with systematic uncertainties of a factor of around five on individual measurements (see ~\citealt{Hainline11, Michalowski12}). Studies of resolved H$\alpha$ and $^{12}$CO emission lines indicate SMGs have dynamical masses of (1--2)\,$\times$10$^{11}$\,$ \Msol$ (\citealt{Swinbank06b,susie12,Bothwell13}), placing an upper limit on the stellar masses, but the number of SMGs with measured dynamical masses is small and the samples inhomogeneous. The spatial clustering of single-dish detected sub-mm sources has been shown to match that expected for the progenitors of local ellipticals \citep{Hickox12}. However, these results are complicated by source blending in the coarse (19$''$) resolution single-dish sub-mm imaging \citep{Hodge13}, whereby the detected sub-mm source comprises of multiple individual SMGs. Finally, while the space densities of SMGs and ellipticals are in agreement, the analysis is highly dependent on the assumed duty cycle of the SMG population (e.g.\ \citealt{Simpson14}). 

The morphologies of SMGs provide an alternative, and potentially powerful, tool for testing any evolutionary connection, since the population of quiescent galaxies at high redshift appear extremely compact in rest--frame optical imaging (half--light radius [$R_{\rm e}$] $\sim 1$\,kpc; e.g.\ \citealt{Daddi05,Zirm07,Toft07,Buitrago08, vandokkum08b,Newman12,Patel13,Krogager13}). Recently, \citet{Chen14} presented Wide Field Camera 3 (WFC3)/{\it HST} imaging of 48 SMGs, finding a median half-light radius of $R_{\rm e}$\,=\,4.4$^{+1.1}_{-0.5}$\,kpc, considerably larger than the quiescent population (see also \citealt{Targett13,Wiklind14}). However, as discussed by \citet{Chen14}, the SMGs in their sample are predominantly disturbed systems, with indications that the intense star formation is triggered by merger activity (see also \citealt{Frayer99,Greve05,Tacconi06,Engel10,Ivison13}). The sizes presented by \citet{Chen14} are thus likely to over-estimate the size of the system at post-coalescence. In addition, the high star formation rates of SMGs, combined with large molecular gas reservoirs, means that they have the potential to at least double their stellar mass during the starburst phase. As such, understanding the spatial distribution of the ongoing star formation is crucial to understanding the stellar distribution of the post starburst galaxy. 

In the local Universe ULIRGs, the proposed analogs of SMGs, appear to be compact (1--2\,kpc diameter) using resolved $^{12}$CO/dust/mid-infrared/radio emission (\citealt{Condon91b, Downes98, Soifer99, Sakamoto08, Rujopakarn11, Ueda14}). At the typical redshift of SMGs ($z$\,=\,2.5) it has only been possible to resolve the dust emission in a small number of the brightest sources. \citet{Younger10} present observations of two bright SMGs ($S_{870}$\,=\,13\,mJy \& 18\,mJy) using the Sub-millimeter Array (SMA). Both of these SMGs appear resolved in the SMA data and have a FWHM of $0.6 \pm 0.2''$; at the typical redshifts of SMGs this corresponds to a physical size of $4$--$5$\,kpc. More recently~\citet{Hodge13} obtained interferometric follow-up observations, with the Atacama Large Millimeter Array (ALMA), of single-dish identified sub-mm sources. \citet{Hodge13} identified 99 SMGs in the $1.6''$ resolution ALMA maps, but found that only one SMG is resolved with a FWHM of $9$\,kpc. The remaining SMGs are unresolved, with sizes $< 10$\,kpc \citep{Hodge13}.

%
% Figure1- Thumbnails of dust emission on K-band image
%
\begin{figure*}
   \centerline{ \psfig{figure= 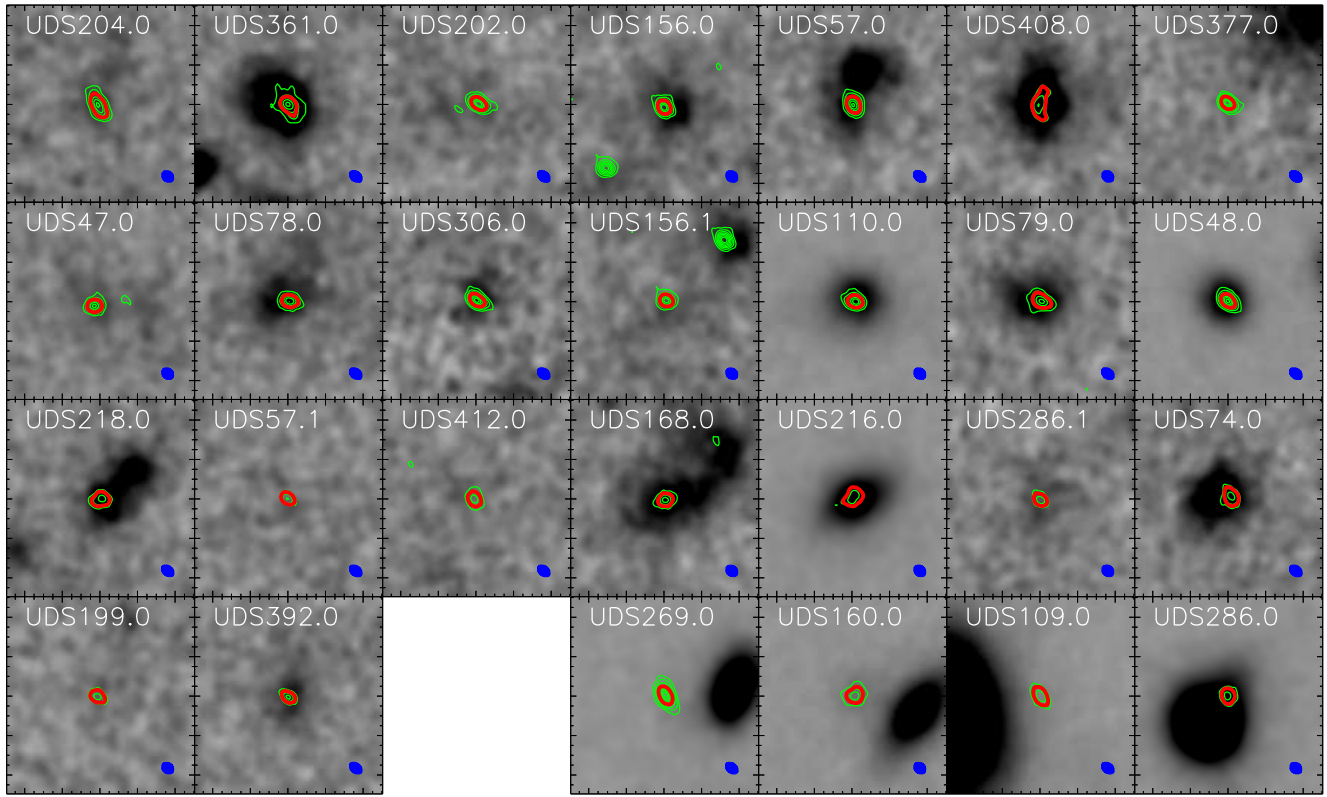,width=0.99\textwidth}}
\caption{ Grayscale K--band images of the 27 bright (SNR\,$>$\,$10$\,$\sigma$) ALMA-identified SMGs in our sample. The images are in order of decreasing 870\,$\mu$m flux density, except for the final four panels, which are classed as potentially lensed SMGs (images are separated by a blank panel). Each panel is 5$'' \times$\,5$''$ and we contour the ALMA maps over the images of the galaxies. The green contours on each image represent ALMA 870\,$\mu$m emission at 4, 8, 12,\,....\,$\times\sigma$, and a single red contour represents where the ALMA 870\,$\mu$m surface brightness is 50\,\pc\, of the peak value; for an ideal point source this contour should be identical to the size of the ALMA beam FWHM (bottom right of each panel). We note that the red contour appears more extended than the beam, indicating that these SMGs are resolved in the 0.3$''$ resolution ALMA data. Overall, 15\,\pc\ of the SMGs are not detected in the K--band imaging (5\,$\sigma$ detection limit 25.0\,mag).
}
 \label{fig:thumbs}
\end{figure*}

Limited studies in the radio and CO also hint that SMGs have sizes of a few kpc in diameter (see~\citet{Menendez09} for a discussion of the mid--infrared sizes of SMGs). However these studies suffer from two limitations. In $^{12}$CO the resolution achieved using facilities such as the Plateau de Bure Interferometer (PdBI) or SMA is barely sufficient to resolve the emission to $\lsim 3$--4\,\,kpc, and studies indicate the SMGs are either unresolved or marginally resolved at this resolution (e.g.\ \citealt{Tacconi06,Engel10}). These $^{12}$CO observations are also typically carried out in the higher-$J$ transitions, which trace the denser and warmer gas, not necessarily reflecting the full extent of the gas reservoir, or star formation activity. Indeed there is evidence from spatially-resolved $^{12}$CO\,($J$\,=\,1--0) observations of a small sample of SMGs that the cool gas extents of these systems are considerably larger than claimed from high-$J$ observations (e.g.\ \citealt{Ivison11EVLA, Riechers11b, Hodge13b}). In contrast to studies of $^{12}$CO emission, radio observations provide the resolution required to resolve SMGs at 1.4\,GHz with eMERLIN (e.g.\ \citealt{Biggs08}) or at higher frequencies with the Karl G. Jansky Very Large Array (VLA). However, such studies in the radio typically rely on the far--infrared radio correlation to identify the sub-mm sources. The identification procedure is inherently probabilistic, and as shown by ~\citet{Hodge13} the reliability and completeness of the identifications is 80\,\pc\ and 50\,pc\, respectively . In addition, the form of the spatially--resolved far-infrared radio correlation is still debated in the local Universe, owing to the potential diffusion/leakage of the cosmic rays and the influence of magnetic fields on the resulting emission \citep{Bicay90, Murphy06, Murphy08}. Hence translating radio sizes into star formation extents is uncertain, especially when extrapolating the results of local studies to SMGs at $z\gsim$\,2.

%
% Figure3- Amp vs UVdistance plots
%
\begin{figure*}
   \centerline{ \psfig{figure=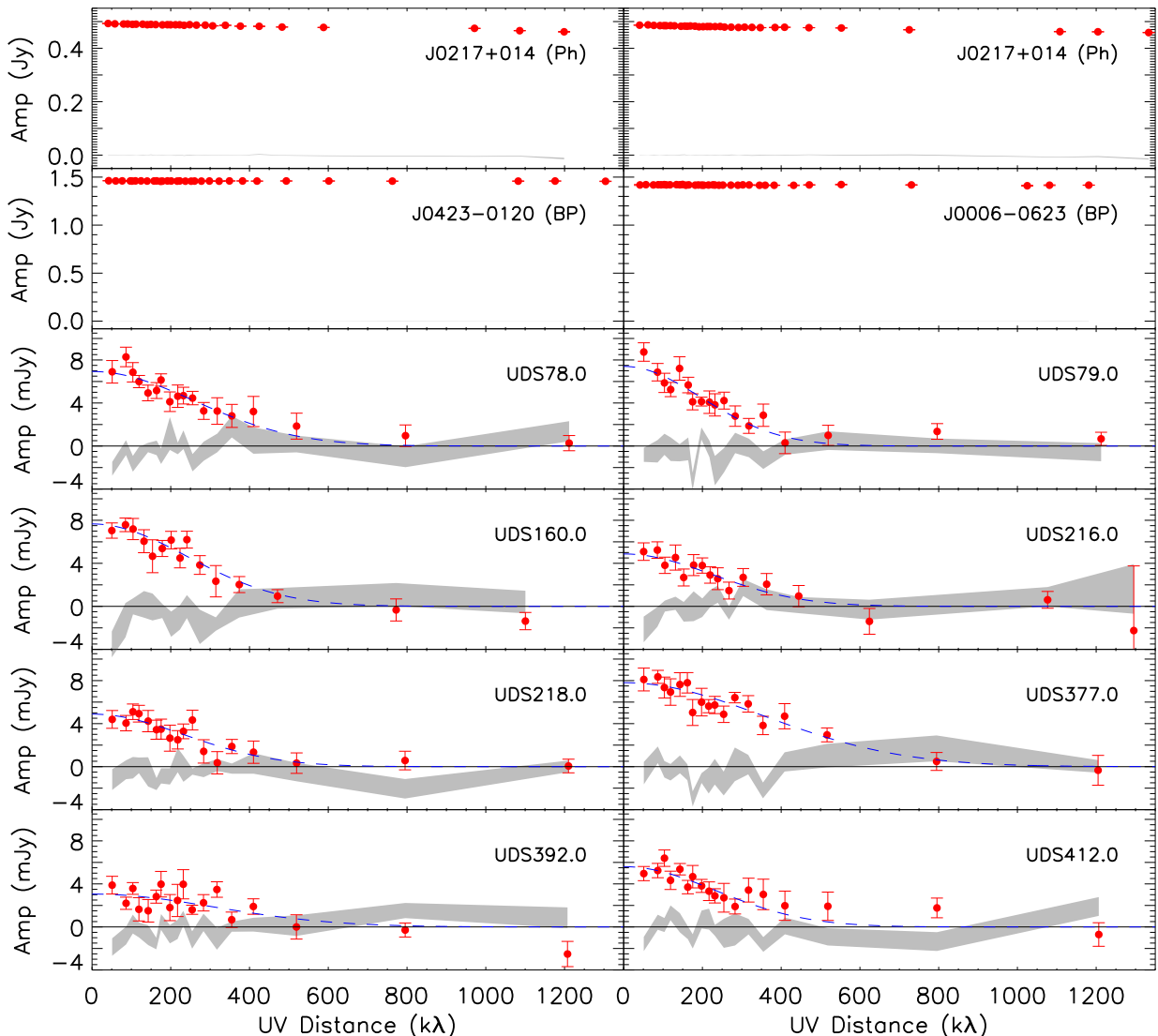,width=0.95\textwidth}}
\caption{Components of the complex visibility versus $uv$-distance for the phase  (Ph) and bandpass (BP) calibrators from both measurement sets (top), and an example eight SMGs from our sample (lower panels). The real components of the complex visibilities are plotted as data points, while the 1-$\sigma$ range of the imaginary components is shown as a grey shaded region. Both the real and imaginary components are plotted on the same scale on the left axis. The amplitudes of both sets of calibrators are relatively flat with $uv$-distance, indicating that they are unresolved point sources, although we note a marginal drop in the flux of the phase calibrator, which suggests it is weakly resolved. In contrast the amplitudes for seven of the SMGs decline strongly with $uv$-distance, confirming our conclusion that these sources are resolved in the sub--mm imaging at 0.3$''$ resolution. We plot on each SMG a dashed line representing the best-fitting Gaussian to the amplitudes. We note that our measurements of the size and flux density of these SMGs in the image plane are consistent with the Gaussian fits to the amplitudes (the median ratio in size is FWHM$^{\rm uv}$\,/\,FWHM$^{\rm image} =  $\,0.9\,$\pm$\,0.2).
}

 \label{fig:ampuv}
\end{figure*}

In this paper we present the results of ALMA 870\,$\mu$m observations of 30 bright sub-millimeter sources. These SMGs are selected at 850-$\mu$m from the SCUBA-2~\citep{Holland13} Cosmology Legacy Survey of the UKIDSS Ultra Deep Survey (UDS) field, and were mapped at an angular resolution of 0.3$''$ FWHM using ALMA. In this work we focus on the sizes and morphologies of these SMGs; the catalogue and number counts will be presented in Simpson et al.\ (in prep). We discuss in \S 2 our sample selection, and the ALMA 870\,$\mu$m observations and their reduction.  In \S 3 we present the sizes of the resolved dust emission in the SMGs, and in \S 4 we present a comparison of these typical SMGs to similar high redshift sources and to local U/LIRGs. We give our conclusions in \S 5. Throughout the paper, we adopt a cosmology with $\Omega_{\Lambda}$\,=\,0.73, $\Omega_{\rm m}$\,=\,0.27, and $H_{\rm  0}$\,=\,71\,km\,s$^{-1}$\,Mpc$^{-1}$, in which an angular size of 1$''$ corresponds to 8.5--7.5\,kpc at $z \sim $\,1.5--3.5. Unless otherwise stated, error estimates are from a bootstrap analysis.

\section{Observations and Analysis}
\label{sec:observations}

%
% Figure3- Residuals
%
\begin{figure*}
   \centerline{ \psfig{figure=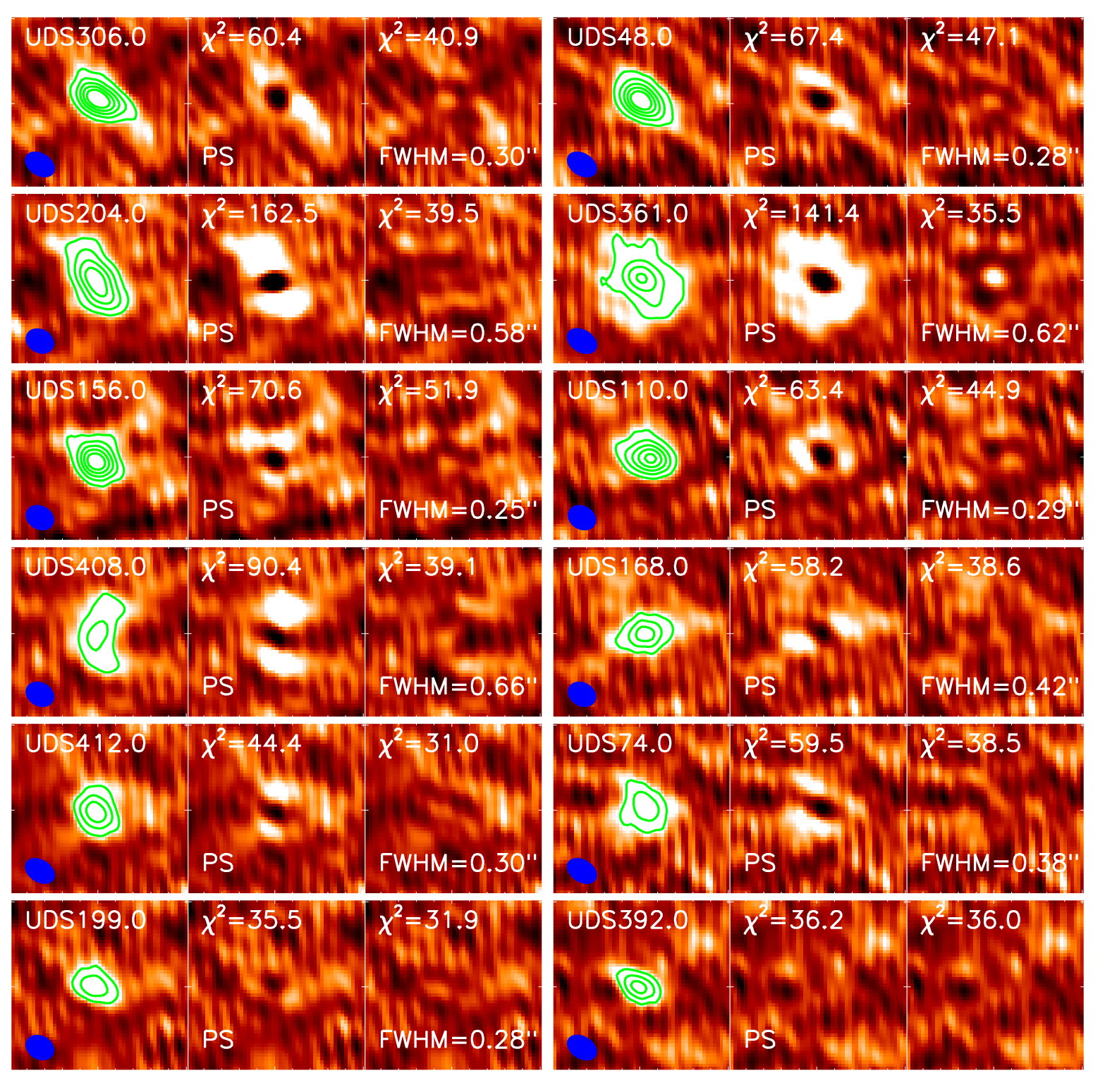,width=0.95\textwidth}}
\caption{ Examples showing 2$''$\,$\times$\,2$''$ images of our high resolution 870\,$\mu$m maps (0.3$''$; left) alongside the residuals from fitting a point source model (PS; middle of each set of three panels) and an elliptical Gaussian model (resolved; right) for 12 example SMGs in our sample. The green contours represent 870\,$\mu$m emission at 4, 8, 12,\,....\,$\times\sigma$, and the color--scale in each image is clipped at $\pm 3$\,$\sigma$. The ALMA beam is shown in the bottom left of the left--hand column. The SMGs presented here are chosen such that they span the full range in detection significance for the sample, and are a fair representation of the data quality of our ALMA maps. In eleven of these images there are significant residuals when fitting a point source model to the 870\,$\mu$m emission, indicating that these sources are resolved in our ALMA maps. In contrast we highlight UDS392.0, which is well--described by a point source model and is classed as unresolved in our analysis. For the full sample of 23 SMGs the median difference in $\chi^2$ between the best-fit extended and point-source model is $\Delta$\,$\chi^{2}$\,=\,20\,$\pm$\,2. We find that 22\,/\,23 SMGs are resolved in our 0.3$''$ resolution imaging, and derive a median angular size for the sample of 0.30\,$\pm$\,0.04$''$ (deconvolved FWHM of the major axis). 
} 
 \label{fig:residuals}
\end{figure*}

The observations discussed here targeted 30 870\,$\mu$m sources lying in the UKIDSS UDS field.  These sources were selected from wide-field 850\,$\mu$m  observations of the UDS field taken as part of the SCUBA--2 Cosmology Legacy Survey (S2CLS) program with the SCUBA-2 camera on the James Clerk Maxwell Telescope (JCMT).  The current SCUBA-2 observations reach a typical depth of $\sigma_{850}$\,=\,1.2\,mJy across a 0.8\,$\times$\,0.8\,degree$^2$ field, and have an angular resolution of 14.8$''$ FWHM. From an early version of these observations, with $\sigma_{850}$\,=\,2\,mJy, we selected a sample of 30 of the brightest sub-millimeter sources in the field, detected at $>$\,4$\sigma$ significance and hence having 850$\mu$m flux densities of 8\,mJy--16\,mJy.

\subsection{ALMA data}
\label{sec:ALMA} 
The data reduction and source extraction from our ALMA data is described in detail in Simpson et al.\ (in prep.). Here we give a brief description. The ALMA data were taken on 2013 November 1, as part of the Cycle-1 project 2012.1.00090.S.  We observed all 30 sub-millimeter sources with ALMA, using 7.5-GHz of bandwidth centered at 344\,GHz (870\,$\mu$m; Band 7); the same frequency as the original SCUBA-2 observations. We used a ``single continuum'' correlator setup with four basebands of 128 dual-polarization channels each. 

The phase centers for the  ALMA pointings are the centroid position of the sub-millimeter source from the early SCUBA-2 map. The ALMA primary beam at this frequency is $17.3''$ FWHM,  larger than the 14.8$''$ FWHM of the SCUBA-2 beam and so is sufficient to recover all of the SMGs contributing to the single-dish sub-millimeter emission. The array configuration for our observations was such that the 26 ALMA 12--m antennae employed had a maximum baseline of 1250\,m, and a median baseline of 200\,m. This is in fact more extended than our requested compact C32--1 configuration and as a result yields a synthesized beam of 0.35$''\times$\,0.25$''$ using Briggs weighting (robust parameter = 0.5). The maximum angular scale that our observations are sensitive to is 5$''$, which as we show in \S\,\ref{sec:Fitting} is an order of magnitude larger than the FWHM of the sources in our sample.

The observations of our 30 targets were split into two 15-target blocks. Each block comprises seven or eight sub-blocks of 30\,s observations
of 10 targets, with the targets each observed five times. The targets were randomly assigned to different sub-blocks; i.e.\ one source might be observed in sub-blocks 1,3,5,6,7, and  another in 1,2,3,4,6. Each block has a full set of calibration observations and each 5-min sub-block is separated by a 90\,s phase calibration observation and a 30\,s atmospheric calibration (taken on the phase calibrator).  Hence, for each map we obtained a 150\,s integration. Absolute flux calibration was derived from J\,0238\,+\,166 and we used observations of the secondary phase calibrator J\,0217\,+\,014, for phase referencing. The total integration time for the project was 2.6\,hrs. 
 
The data were imaged using the {\sc Common Astronomy Software Application} ({\sc casa} version 4.2.1). As detailed in Simpson et al.\ (in prep) the $uv$-data were Fourier transformed to create a ``dirty'' image.  We then applied the same procedure adopted by Hodge et al.\ (2013): a tight clean box was placed around all $>$\,5\,$\sigma$ emission in each map and these were cleaned down to a depth of 1.5\,$\sigma$. The final cleaned maps have a median angular resolution of 0.35$''\times$\,0.25$''$ (P.A. $\sim$ 55\,deg), using Briggs weighting (robust parameter = 0.5), and a median rms of $\sigma_{870} = $\,0.21\,mJy\,beam$^{-1}$ (with a range from 0.19--0.24\,mJy\,beam$^{-1}$). We inspect our maps and find that we do not detect any sources at sufficiently high SNR to reliably self--calibrate. 

To construct the master catalog, and ensure that the extended flux from the SMGs is recovered, we repeated the imaging procedure described above, but using natural weighting, and applying a Gaussian taper to the data in the $uv$-plane. Applying a Gaussian taper means that a lower weighting is given to visibilities at large distances in the $uv$-plane, producing a map with a larger synthesized beam, hence with less emission resolved out, at the expense of higher noise. The resulting maps have a median angular resolution of 0.8\,$'' \times $\,0.65$''$ , and median rms of $\sigma_{870} = $\,0.26\,mJy\,beam$^{-1}$. Hence two sets of maps were created; ``detection''  maps with a synthesised beam of $\sim $\,0.8$''$ FWHM and ``high-resolution'' maps at $\sim$\,0.3$''$ FWHM.  

To construct our catalog, we identify sources within the ALMA primary beam FWHM that are detected at $>$\,4\,$\sigma$ in the 0.8$''$ FWHM ``detection'' maps, and extract both the peak flux density and the total flux density in a 0.8$''$ radius aperture for each SMG. We search for sources outside the ALMA primary beam FWHM, but do not find any statistically significant detections (see Simpson et al.\, in prep). In total we identify 52 SMGs above 4\,$\sigma$ in the 30 ALMA maps, at 0.8$''$ FWHM resolution. These SMGs have a range of 870\,$\mu$m flux density of 1--14\,mJy and we recover the single--dish SCUBA-2 with a median ratio of $S^{\rm SCUBA2}$\,/\,$S^{\rm ALMA}$\,=\,1.04\,$\pm$0.05. As the analysis presented in this paper is focused on the size distribution of SMGs, we cut our SMG sample at a higher significance limit to provide sufficiently high signal-to-noise ratios (SNR) for the SMGs to reliably measure their sizes. As we show below, this corresponds to SNR\,$\geq$\,10 ($S_{870}$\,$\gsim$\,4\,mJy) at 0.8$''$ resolution, which reduces our sample to 27 SMGs. In order to check the measured fluxes of these sources we also used the {\sc imfit} routine in {\sc casa} to model each SMG and find good agreement between the model and aperture derived flux densities ($S^{\rm Model}$\,/\,$S^{\rm Aperture}$\,=\,1.02\,$\pm$\,0.01). Where appropriate in the following analysis, we will also use the average properties of the remaining 25 SMGs with $S_{870}$\,$\lsim$\,4\,mJy to test for trends with flux density across the whole sample. 

Given the relatively bright $\gsim$\,8\,mJy flux limit used to select our target sample, we need to be aware of the potential influence of gravitational lensing (e.g.\ \citealt{Blain96a,Chapman02})\,\footnote{ Lensing models predict that our sample contains 1--2 SMGs that are lensed with an amplification factor $>$\,2 (e.g. \citealt{Paciga09}. However we note that the prediction is based on the~\citet{Chapman05} redshift distribution for SMGs, which does not contain the high redshift tail of sources seen in other SMG redshift distributions (e.g. \citealt{Yun12,Smolcic12,Weiss13,Simpson14}). Hence the prediction for the number density of lensed sources is likely to be a lower limit.}.  Indeed, we identify four examples of potential gravitationally-lensed sources in our bright SMG sample: UDS\,109.0, UDS\,160.0, UDS\,269.0, and UDS\,286.0.  All of these sources appear to be close to, but spatially offset from, galaxies at $z\ls$\,1 (Figure~1). If these sources are lensed then their apparent sizes need to be corrected for lens amplification. However, they all appear singly-imaged (with no sign of multiple images at our sensitive limits) and as we have no precise estimates of their redshifts or the masses of the foreground lenses, it is impossible to reliably determine this correction. We therefore highlight these four SMGs in Figure~1, and note that the derived sizes for these sources are likely to be over--estimated. When discussing the properties of our sample we do not include the potentially lensed sources in the analysis, and so our final sample consists of 23 SMGs detected at SNR\,$>$\,$10$\,$\sigma$

\subsection{Robustness of imaging}
As we show in \S\,\ref{sec:Fitting} most of the SMGs in our sample are resolved in the 0.3$''$ resolution ALMA maps. However, we now perform two tests to ensure that an error in the calibration of the raw data does not drive our conclusion that the sources are resolved.

The observing strategy for our targets was such that each SMG was observed five separate times through the observing block, with phase calibration observations between each repeat observation. First, we test for variations in the source size as a function of time, as might be expected if the phase solution applied to the data does not correctly model fluctuations. To do so, we separately image each repeat observation of UDS\,204.0, the brightest SMG in our sample (the integration time is sufficient to detect the source at 8--10\,$\sigma$ in an individual scan). The SMG appears resolved in all five images, with an intrinsic source size in the range $0.52''$--$0.63''$, and all sizes consistent within the associated 1--$\sigma$ uncertainties ($\sim$\,0.05$''$).

Secondly, we re-classify an observation of the phase calibrator as a science target observation in the observing sequence and repeat the calibration of the data set. If an error in the calibration of the raw data, due to phase variations, are causing the SMGs in our sample to appear resolved then the phase calibrator will also appear resolved in this scan. We image this phase calibrator scan in the same manner as the SMG observations and model the emission using the {\sc imfit} routine. In the individual scan the phase calibrator has an intrinsic size of FWHM\,=\,0.06\,$\pm$\,0.01$''$, which is marginally higher that the intrinsic size measured in an image combining all scans (FWHM\,=\,0.03\,$\pm$\,0.01$''$). However, by removing a phase calibrator scan we have introduced a ten minute gap in our calibration observations, which is double that used throughout the observations. Hence any difference in the size of the phase calibrator should be classed as an upper limit on the uncertainty due to errors in the phase calibration of the data. Taken together these tests indicate that the SMGs in our sample do not appear resolved due to errors in the calibration of the raw data. 

In our analysis in \S\,\ref{sec:Fitting} we use the {\sc imfit} routine in {\sc casa} to fit an elliptical Gaussian model (convolved with the synthesised beam) for the 870\,$\mu$m emission from each SMG. However, before applying this approach to the bright SMGs in our sample, we now test the reliability of the fits as a function of  SNR using simple simulated galaxies. To do so, we create 20,000 elliptical Gaussian model sources with a uniform distribution of peak SNR from 4 to 30 (covering the range of SMGs in our full sample) and a major-axis size distribution that is uniform from 0 to 1$''$. The model sources are distributed uniformly in SNR at 0.8$''$ resolution, reflecting the selection function of our full SMG catalog. In order to use realistic noise maps we add these models to a randomly chosen position in one of the residual (source-subtracted) ALMA maps.  

We use {\sc casa} to fit an elliptical Gaussian to each injected model source and derive the best-fit size for each.  We note that the {\sc imfit} routine can return a point source solution if this is the best model. For each model parameter we calculate the fractional offset between the input model value and the recovered value. As expected we find that the precision of the recovered parameters is a function of SNR. To ensure that the sizes we derive for the SMGs are reliable we define a selection limit that the 1-$\sigma$ spread in (\,FWHM$^{\rm Model} - $FWHM$^{\rm True})$\,/\,FWHM$^{\rm True} < $\,0.3,  i.e.\ 68\% of the model sources are recovered with a fractional error less than 30\%. We find that this requirement is met for sources that are detected at a SNR\,$\geq$\,10, which is the justification for our use of this limit to define the bright SMG sample analysed in this paper.  

These simulations also indicate a small bias in the recovered size for sources with SNR between 10 to 30\,$\sigma$ of (\,FWHM$^{\rm Model} - $FWHM$^{\rm True})$\,/\,FWHM$^{\rm True}$\,=\,0.018\,$\pm$\,0.002 (1--$\sigma$ spread of $\pm 0.14$), which we have not corrected for in the following analysis. The simulations also allow us to investigate the resolution limit of our maps, by determining the true size of a model source for which the {\sc imfit} routine returns a point source best-fit solution. For SMGs with SNR\,$=$\,10--30 we find that 90\% of the model sources that are best fit by point source models using {\sc imfit} actually have a ``true'' size $\leq$\,0.18$''$ (i.e.\ smaller than half the size of the beam major axis). Hence, we can be confident that any SMGs that have a point source best-fit model have a size $\leq$\,0.18$''$, and we adopt this as an upper limit for the size of the unresolved SMGs.

%
% Figure4 - deconvolved sizes vs S870 etc
%
\begin{figure*}
   \centerline{ \psfig{figure=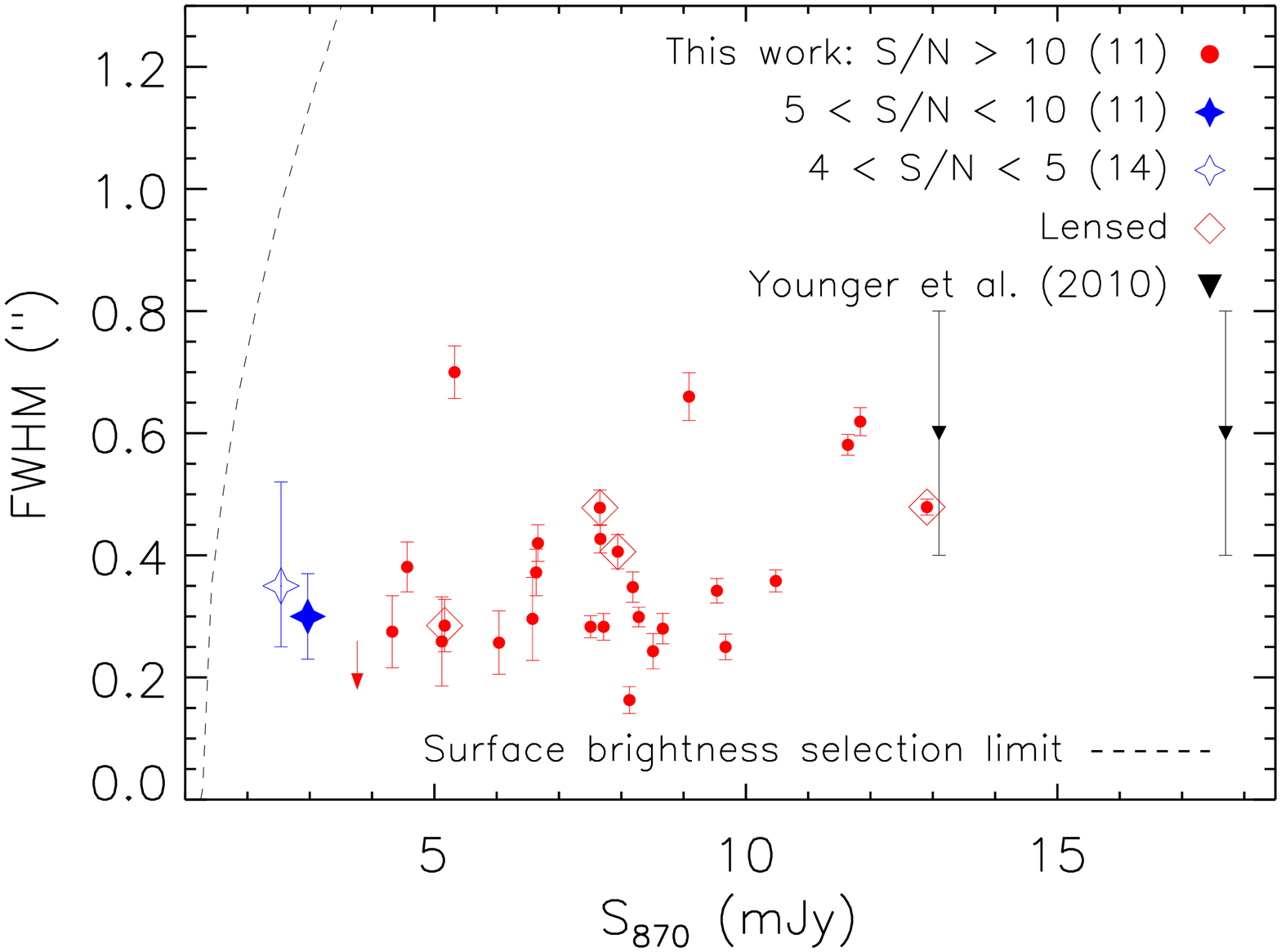,angle=90,width=0.46\textwidth} }
   \vspace{0.2cm}
   \centerline{   \psfig{figure=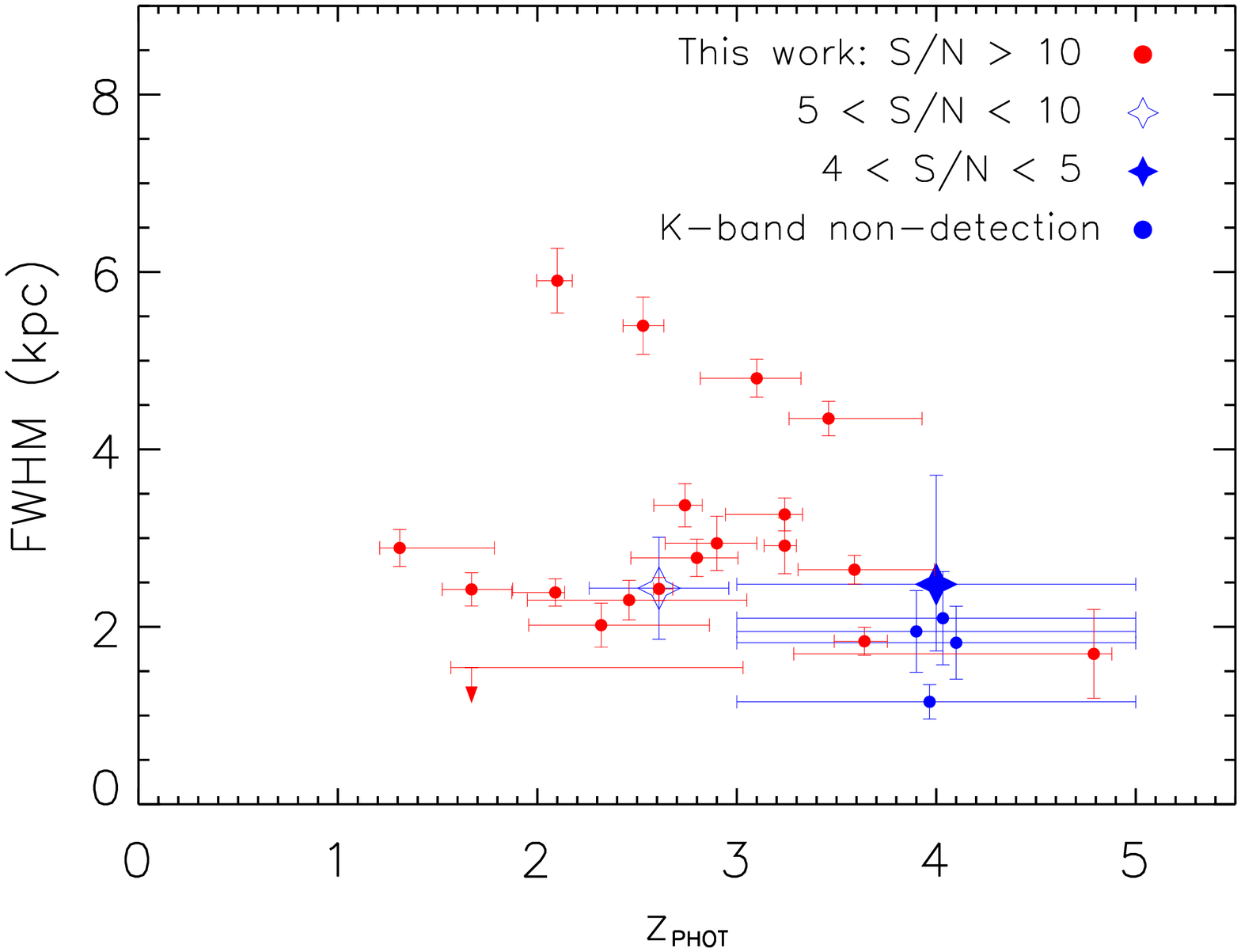,angle=90,width=0.46\textwidth}
   \hspace{-1.6cm}
   \psfig{figure=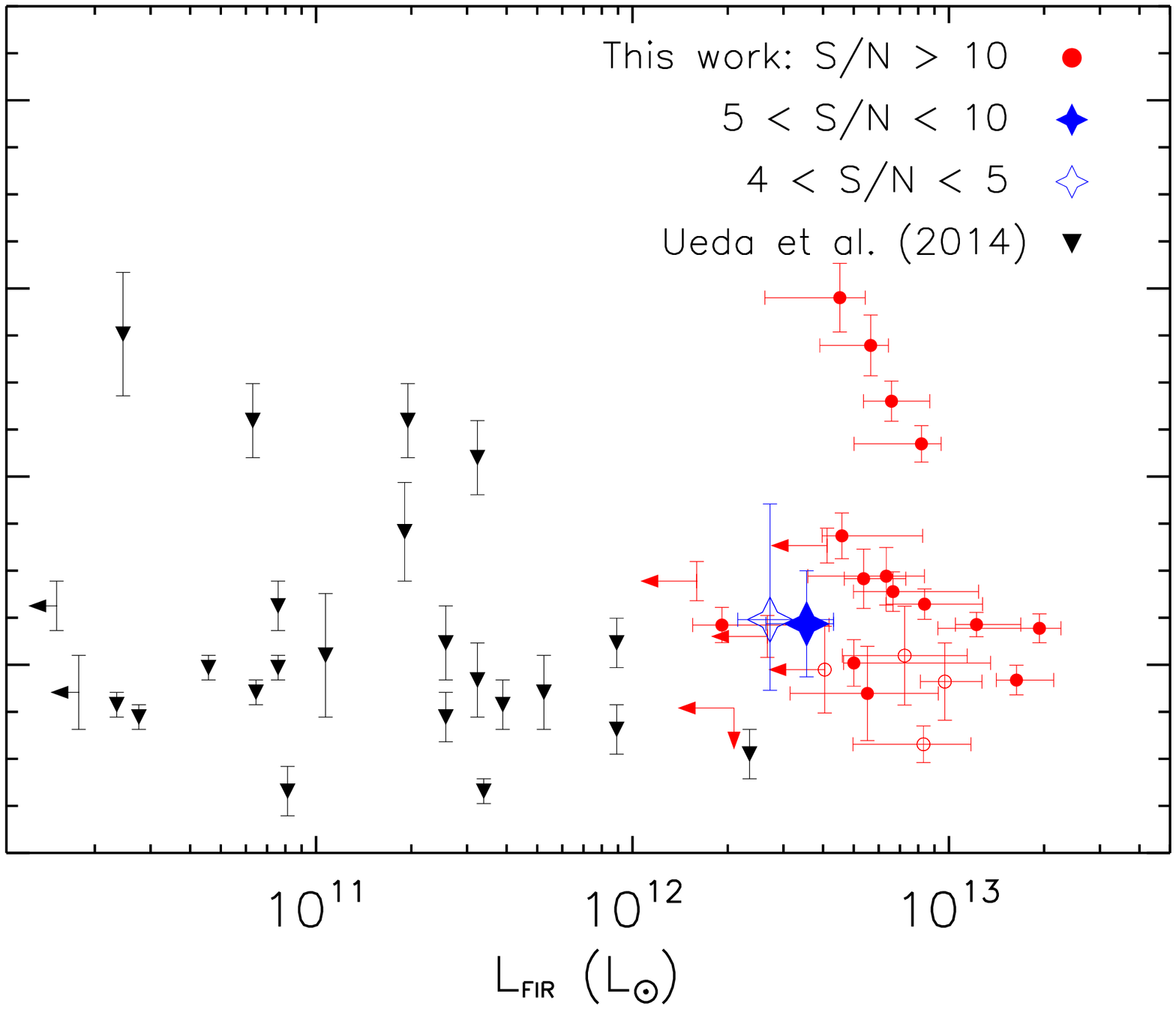,angle=90,width=0.46\textwidth}}

\caption{ {\bf{Top:}} Angular size distribution of the 870\,$\mu$m emission from SMGs as a function of their 870-$\mu$m flux density. A dashed line represents our surface brightness selection limit. The 23 SMGs detected at $\geq$\,10\,$\sigma$, have a median intrinsic size of 0.30\,$\pm$\,0.04$''$ (deconvolved FWHM of the major axis). We stack the 870\,$\mu$m emission from SMGs detected at 5--10\,$\sigma$ and 4--5\,$\sigma$, and show the size derived from each stack. We find that these 870\,$\mu$m--faint SMGs have sizes that are on average consistent with the brighter SMG distribution. We highlight four SMGs that are potentially lensed sources, but note that these are not included in our analysis. For comparison we show the sizes of two bright SMGs measured from observations with the SMA \citep{Younger10}. {\bf{Bottom left:}} Physical size of SMGs as function of redshift. Four SMGs in our sample do not have a photometric redshift, and we fix the redshift of these sources at $z = 4 \pm 1$ (the redshifts are offset in the plot). The potentially lensed SMGs are not shown on this figure. The median physical size of the 23 SMGs in the sample is FWHM\,$= 2.4 \pm 0.2$\,kpc, and we do not find a trend in the physical size with redshift. From our stacking analysis we measure an average physical size for the SMGs detected at SNR\,$=5$--10\,$\sigma$ of $2.4 \pm 0.6$\,kpc, consistent with our results for the $870$\,$\mu$m--brighter SMGs. We also show the results of stacking the 14 SMGs detected at SNR\,$ = 4$--5\,$\sigma$, but note that 8/14 sources in the stack do not have a photometric redshift. {\bf{Bottom right:}} Physical size of the SMGs in our sample as a function of FIR-luminosity. The 23 SMGs shown have a median FIR luminosity of $L_{\rm FIR} = $\,(5.7\,$\pm$\,0.7)$\times$10$^{12}$\,L$_\odot$, and a median physical size of FWHM\,$= 2.4 \pm 0.2$\,kpc. We do not find a trend between $L_{\rm FIR}$ and the physical size of the SMGs.  For comparison we also show the sizes of local ``merger--remnants'' measured from interferometric observations of the $^{12}$CO\,($J$\,=\,1--0\,/\,2--1\,/\,3--2) molecular emission line. These local galaxies have a median size of $1.9 \pm 0.2$\,kpc, and are marginally smaller than the SMGs presented here. As the ``low''--$J$ $^{12}$CO typically exceeds the size of the star forming region, the star formation in SMGs appears to be considerably more extended than in local U/LIRGs.
}
\label{fig:size870}
\end{figure*}

\subsection{Multi--wavelength properties}

During our analysis we use archival multi--wavelength information on this well-studied field. An extensive analysis of the multi--wavelength properties of the SMG sample  will be presented in Ma et al.\ (in prep), but for the purposes of this paper we use two pieces of information from that work:  constraints on the likely redshifts of the SMGs using photometric redshifts; and estimates of their far-infrared luminosities.  

First, we use the photometric redshift catalog of this field produced by the UKIDSS UDS team, which is based on a $K$-band selected sample of sources. To briefly summarise, photometric redshifts are determined for each field source by fitting template spectral energy distributions (SEDs) to the observed $UBVRi'z'JHK$ and IRAC 3.6 and 4.5\,$\mu$m photometry, using the public SED fitting code {\sc eazy} \citep{Bramer08}. Excellent agreement is found between the photometric and spectroscopic redshifts for 2146 sources in the field, with a median dispersion of $(z_{\rm   phot} - z_{\rm spec})$\,/\,$(1+z_{\rm spec}) = 0.03$ (see \citealt{Hartley13,Mortlock13}). 

Matching the SMG catalog to the photometric redshift sample finds 19 matches within 1$''$ of the 23 bright SMGs, which leaves four of the SMGs without redshift information. We also searched for IRAC counterparts for the four SMGs without photometric redshifts and find that three have detections in the IRAC imaging; however, the limited number of photometric bands available for all of these sources mean it is impossible to derive precise photometric redshifts. \citet{Simpson14} argue that these NIR--faint SMGs typically lie at higher redshifts than the optical/NIR--brighter SMGs, and so we adopt their approach and assign these sources a redshift of $z = $\,4\,$\pm$\,1.

To derive the FIR luminosities of the SMGs we exploit the {\it Herschel} SPIRE imaging~\citep{Pilbratt10,Griffin10} of the field at  250, 350 and 500\,$\mu$m, along with the precise position of the ALMA source to deblend and extract flux densities for the SMGs. Following the approach of \citet{Swinbank13}, they use a prior catalog of sources detected at 24\,$\mu$m or 1.4\,GHz, along with the ALMA source positions, to model the sources contributing to the map flux in the vicinity of each SMG. Having extracted flux densities or limits in the three SPIRE bands, along with the  ALMA 870\,$\mu$m flux density, they fit a library of model spectral energy distribution (SED) templates (see \citealt{Swinbank13}), to the FIR-photometry and determine the best-fit template. We integrate the best-fit SED from 8 to 1000\,$\mu$m to derive the total far-infrared luminosity of each SMG. If an SMG is undetected in the {\it Herschel} 250, 350 and 500\,$\mu$m imaging then the far-infrared luminosity is treated as an upper limit. We measure a median far-infrared luminosity for the 23 SMGs in our sample of $L_{\rm FIR} = $\,(5.7\,$\pm$\, 0.7)\,$\times$10$^{12}$\,L$_\odot$, and a median dust temperature of $T_{\rm d} = $\,32\,$\pm$\,3\,K. We find that 5\,/\,11 and 10\,/\,14 SMGs detected at SNR\,=\,5--10\,$\sigma$ and SNR\,=\,4--5\,$\sigma$, respectively, are undetected in the {\it Herschel} 250, 350 and 500\,$\mu$m imaging. We stack the {\it Herschel} SPIRE maps for these subsets and derive an average $L_{\rm FIR} = $\,$3.6^{+0.8}_{-0.9}$\,$\times$10$^{12}$\,L$_\odot$ and $L_{\rm FIR} = $\,$2.7^{+1.6}_{-0.6}$\,$\times$10$^{12}$\,L$_\odot$ for the SNR\,=\,5--10\,$\sigma$ and SNR\,=\,4--5\,$\sigma$ subsets, respectively.

\section{Results}

\subsection{Resolved  dust emission}
\label{sec:Fitting} 

We show the ALMA maps with a 0.3$''$ FWHM resolution synthesized beam for the 27 SNR\,$\geq$\,10 SMGs in Figure~1. These are presented as contours overlaid on $K$--band grayscale images, to allow the reader to make a qualitative comparison of the size and shape of the contour corresponding to the half-peak-flux level and the corresponding size for the synthesized beam.  We see evidence in a majority of the SMGs that the sub-millimeter emission appears to be more extended than would be expected for an unresolved source with the same peak flux, with several examples also showing structure on smaller scales (e.g.\ UDS\,74.0, UDS\,216.0, UDS\,218.0, UDS\,361.0, UDS\,408.0).   

To quantitatively test if the SMGs are resolved in our data we initially perform two non-parametric tests. First we compare the peak flux of each source in the 0.8$''$ FWHM observations to the higher resolution 0.3$''$ FWHM imaging (note that the maps are calibrated in Jy\,beam$^{-1}$). The peak flux of the SMGs is lower in the 0.3$''$ resolution maps, with a median ratio of $S^{0.3}_{\rm pk}$\,/\,$S^{0.8}_{\rm pk} = $\,0.65\,$ \pm$\,0.02, and a 1-$\sigma$ range of $\pm$\,0.04. A drop in the peak flux density between the sources imaged at different resolutions indicates that flux from each SMG is indeed more resolved in the higher-resolution imaging. We note that a 40\% reduction in the peak flux between the 0.3$''$ and 0.8$''$ resolution imaging corresponds to an intrinsic source size of $\sim$\,0.3$''$.  

As a second test we compare the total flux density in an aperture for each source to the peak flux density at both 0.3$''$ and 0.8$''$ FWHM resolution. We convert each map into units of Jy\,pixel$^{-1}$ and measure the ``total'' flux density in an 0.8$''$ radius aperture. The ratio of the peak-to-total flux density is 0.50\,$\pm$\,0.03 at 0.3$''$ FWHM resolution,  compared to 0.83\,$\pm$\,0.03 at 0.8$''$ FWHM. This supports the conclusion that the bulk of the bright SMGs are resolved by ALMA at 870\,$\mu$m with a 0.3$''$ FWHM synthesised beam, and suggests that the average angular size of the population is 0.3--0.4$''$. We note that the aperture fluxes of the SMGs measured in the 0.3$''$ and 0.8$''$ imaging are in good agreement, with a median ratio of $S^{0.3}_{\rm Aper}$\,/\,$S^{0.8}_{\rm Aper} = $\,1.02\,$\pm$\,0.03.

We next investigate whether the SMGs appear resolved in the $uv$-plane, rather than in the image plane, compared to the calibrator sources used for our observations (which are expected to be unresolved). For each source we align the phase centre of the map to the position of the SMG or calibrator source, and extract the amplitudes for each source as a function of $uv$-distance. The amplitudes represent the observed flux of the source on different angular scales in the image plane, with the longest baseline providing information on the smallest angular scales that our observations are sensitive to. For an ideal point source the amplitudes should be constant with $uv$-distance.  As Figure~2 demonstrates we do indeed recover an effectively flat distribution for both the phase and bandpass calibrators. 

In Figure~2 the amplitudes as a function of $uv$-distance for the eight SMGs which are the sole sources detected in their respective ALMA maps (restricting the sample in this way removes any complications due to modelling and subtracting other sources in the primary beam). The amplitudes of seven of the SMGs clearly decrease with increasing $uv$-distance. A single SMG, UDS\,392.0, appears to be only marginally resolved on the longest baselines (our analysis in the image plane identifies this SMG as unresolved, see Fig\ 2 and 3). We fit the amplitudes for each SMG with a single Gaussian component and determine the equivalent size in the $uv$-plane. The model fits in the $uv$ and image planes are consistent, and derive a median size ratio between the image- and $uv$-derived sizes of FWHM$^{\rm uv}$\,/\,FWHM$^{\rm image} = $\,0.9\,$\pm$\,0.2 and a median flux density ratio $S^{\rm image}$\,/\,$S^{uv} = $\,1.1\,$\pm$\,0.1. In 2--3 of the SMGs the flux density does not fall to zero on baselines $\gs $\,600\,k$\lambda$, indicating that there is unresolved flux in these sources on an angular scale of $\ls$\,0.3$''$, but this unresolved component comprises $\ls $\,10\,\pc\ of the total flux density.   

%
% Figure5- Histogram of sizes
%
\begin{figure}
   \centerline{ \psfig{figure=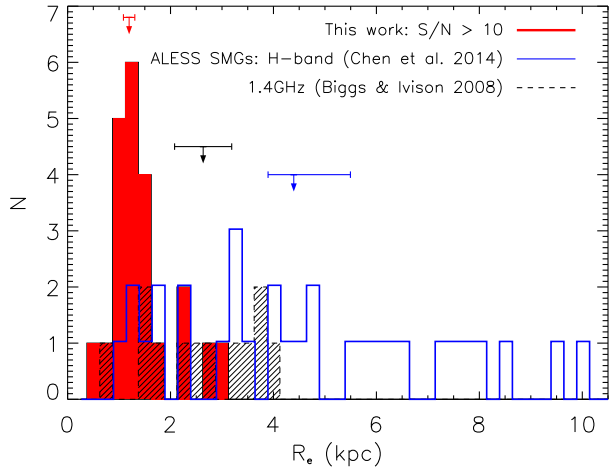,angle=90,width=0.48\textwidth}}
\caption{Comparison of the 870\,$\mu$m sizes of the SMGs in our sample, to the sizes of SMGs measured in 1.4\,GHz\,/\,VLA imaging \citep{Biggs08} and $H$--band/{\it HST} imaging \citep{Chen14}. To allow a fair comparison between the samples we present all of the sizes in terms of an effective radius. The median size of the 12 SMGs identified in the 1.4\,GHz imaging is $R_{\rm e} =2.6 \pm 0.6$\,kpc, approximately double the median $870$\,$\mu$m size of SMGs, $R_{\rm e} = 1.2 \pm 0.1$\,kpc. Convolving the $870$\,$\mu$m emission with a kernel of scale length $1$--$2$\,kpc, which is appropriate for modelling cosmic ray diffusion in star-forming galaxies at low redshift \citep{Murphy08}, increases the median size of the SMGs to $3.8$--$5.2$\,kpc. \citet{Chen14} measure a median size of $R_{\rm e} =$\,4.4$^{+1.1}_{-0.5}$\,kpc from rest--frame optical {\it HST} imaging of 17 interferometrically--identified SMGs at $z$\,=\,$1$--3. The $H$-band sizes of these SMGs are on average 3 times larger than the 870\,$\mu$m sizes presented here, indicating that the obscured star-forming region in SMGs is compact relative to the stellar emission.
}

 \label{fig:hist}
\end{figure}

Given that our non-parametric tests indicate that the SMGs are resolved in our 0.3$''$ imaging, we now chose to fit a more complex model to the sources. Using the {\sc imfit} routine we determine the best-fit elliptical Gaussian model for the 23 SMGs in our bright sample (SNR\,$>$\,10\,$\sigma$). The free parameters of the model are position; peak flux density; major and minor axis; and position angle. The median intrinsic size of the 23 SMGs is 0.30\,$ \pm $\,0.04$''$ (deconvolved FWHM of the major axis), and one SMG is modelled as a point source; as the intrinsic sizes of the sources are comparable to the beam we choose to focus our analysis on the FWHM of the major axis. The median $\chi^{2}$ for the elliptical Gaussian fit is 38.3\,$\pm$\,1.4, and the number of degrees of freedom of the fit is 33 (we define beam to be the size of the resolution element). If we fit each SMG with a point source model then the median difference in $\chi^2$ between the best-fit extended and point-source model is $\Delta$\,$\chi^{2}$\,=\,20\,$\pm$\,2 (see Figure~\ref{fig:residuals})\,\footnote{ We also determined the best-fit Gaussian model for the SMGs in the 0.8$''$ imaging but find that 65\,\pc\ are unresolved, consistent with our results from the higher resolution (0.3$''$) imaging. The resolved sources in the 0.8$''$ imaging have a median FWHM ratio between the two sets of imaging of FWHM$^{\rm 0.8}$\,/\,FWHM$^{\rm 0.3}$\,=\,1.2\,$\pm$\,0.1, however we strongly caution that we have not included upper limits in this calculation and hence the ratio will be biased towards higher values.}.

The top panel of Figure~\ref{fig:size870} shows the sizes of the SMGs as a function of their $870$\,$\mu$m flux density. The dispersion of the 870\,$\mu$m sizes is small and there is no clear trend in the size of the dust emission region with $870$\,$\mu$m flux density. We note however that the two SMG with $S_{870} \sim 12$\,mJy in the sample are around two times larger than the median of the sample. While this may be interpreted as a weak trend in the size of SMGs with 870--$\mu$m flux density we caution against a strong conclusion, given the limited sample size at these fluxes. We also show the sizes of two bright SMGs ($S_{870}$\,=\,13\,mJy and 18\,mJy) presented by \citet{Younger10}. These SMGs were observed with the SMA and both appear marginally resolved in the $uv$--plane, with sizes of $0.6\pm 0.2''$ (deconvolved FWHM). While the size measurements for these sources have large associated uncertainties, they are consistent with the sizes presented here. 

To search for trends in our sample over a larger dynamical range we have also used the fainter SMGs detected in our ALMA maps. While the individual sizes measured for these galaxies have  significant uncertainties, we can attempt to derive a typical size for samples of faint SMGs via stacking (this is reasonable as the synthesised beam does not vary significantly between observations; the beam major and minor axes vary by at most 0.03$''$ and 0.02$''$, respectively across the full sample). We split the fainter 25 SMGs into roughly equal sub-samples detected at SNR\,$=$\,5--10\,$\sigma$ (median $S_{870}$\,=\,3.1\,$\pm$\,0.5\,mJy) and  SNR\,$=$\,4--5\,$\sigma$ (median $S_{870}$\,=\,2.6\,$\pm$\,0.3\,mJy), containing 11 and 14 SMGs, respectively. We stack the 0.3$''$ FWHM maps of these sub-samples and use {\sc imfit} to measure the size of the stacked profile. We derive sizes of 0.30\,$\pm$\,0.07$''$ and 0.35$_{-0.10}^{+0.17}$$''$ for the two sub-samples respectively. The errors on the measured size of each subset are derived by bootstrapping the maps included in the stacks and repeating the analysis. These results indicate that at least on average the 1--4\,mJy SMGs in our full sample have the same size distribution as those with observed 870\,$\mu$m flux densities of 4--15\,mJy sources (see Figure~\ref{fig:size870}).

\section{Discussion}

\subsection{Physical size of SMGs}
\label{sec:trends}
The 23 SMGs detected at $>$\,$10$\,$\sigma$ in our sample have a median angular FWHM of 0.30\,$ \pm $\,0.04$''$. For each SMG we use the photometric redshift to convert the measured angular diameter into an intrinsic physical scale, and derive that the median physical size of the SMGs is $2.4 \pm 0.2$\,kpc (FWHM of the major axis). As shown in Figure~4, the distribution has a narrow interquartile range of 1.8--3.2\,kpc, indicating that the dispersion in the sizes of our sample of SMGs is small. We use the results of our stacking analysis to derive an average physical size of $2.4 \pm 0.6$\,kpc and $2.5_{-0.7}^{+1.2}$\,kpc, at a median redshift of $z = 2.6 \pm 0.4$ and $z = 4 \pm 1$, for the SNR\,$=$\,5--10\,$\sigma$ and SNR\,$=$\,4--5\,$\sigma$ subsets, respectively, indicating no trend in the size of SMGs with redshift. An important caveat of this result is that only one SMG in our sample has a photometric redshift of $z \gsim 3.5$. Four SMGs do not have a photometric redshift and it is likely that these sources lie at $z \sim 4$ (see \citealt{Simpson14}); however the uncertainties on these redshifts are large. At $z$\,=\,4 these SMGs have a distribution of physical FWHM from $<$\,1.3--2.1\,kpc, and are consistent with the lower redshift SMGs in the sample. However, if they instead lie at $z$\,=\,6, then the range of FWHM is $<$\,1.1--1.7\,kpc, and these sources would be smaller than all but one $z < 4$ SMG. 

Recently~\citet{Weiss13} discussed that evolution in the size of SMGs with redshift is a possible explanation for the discrepancy between the redshift distribution of lensed and unlensed SMGs. As 60\pc\, of the SMGs presented by~\citet{Weiss13} lie at $z > 3.5$ we cannot investigate this claim further. However, we note that for the {\it unlensed} SMGs presented here, we do not find any evolution in the sizes of the 870\,$\mu$m emitting region at $z$\,$<$\,3.5.

\subsubsection{Comparison of the multi-wavelength sizes of SMGs}
\label{subsubsec:multisizes}
We now compare the 870\,$\mu$m sizes presented here to the sizes of SMGs measured from observations at different wavelengths. First, we compare our results to a sample of 12 SMGs (median $S_{850}$\,=\,6.8\,$\pm$\,0.6\,mJy) with deconvolved sizes measured from $0.5''$ resolution 1.4\,GHz/{\sc VLA} imaging~\citep{Biggs08}. In their analysis~\citet{Biggs08} fit an elliptical Gaussian model to the 1.4\,GHz emission from each SMG, deriving a median angular size of FWHM\,=\,$0.64 \pm 0.14''$. At the redshifts of these SMGs this angular size corresponds to a median physical size of FWHM\,$= 5.3\pm 1.1$\,kpc. Four SMGs in their sample do not have a redshift and we set these to the median redshift of the sample ($z=2$; consistent with the radio--detected SMGs in~\citealt{Simpson14}). Hence the 1.4\,GHz sizes for the SMGs presented by~\citet{Biggs08} are about two times larger than the 870\,$\mu$m sizes presented here.

It is important to initially note that the radio counterparts to the sub-mm emission presented in \citet{Biggs08} were identified via a probabilistic approach using the far-infrared radio correlation. As shown by~\citet{Hodge13} these identifications are $80$\,\pc\, reliable, and so we might expect that 2--or--3 of the radio sources presented in \citet{Biggs08} are not the true counterpart to the sub-mm emission. However, we do not know if such mis--identifications are likely to be biased towards galaxies with larger, or smaller, sizes at 1.4\,GHz.

Another possible explanation for the discrepancy between the radio and far--infrared sizes of SMGs is that the diffusion length of cosmic rays ($\sim 1$--2\,kpc) is expected to be larger than the diffusion length of far-infrared photons ($\sim 100$\,pc), due to magnetic fields efficiently transporting cosmic rays through the host galaxy. \citet{Murphy08} present a study of 18 local star-forming galaxies with resolved imaging at 70\,$\mu$m and 1.4\,GHz. They show that the radio emission from these galaxies can be modelled as the convolution of the far-infrared emission with an exponentially declining kernel, with an $e$-folding length of the kernel of $1$--$2$\,kpc (see also~\citealt{Bicay90,Murphy06}). We convolve the median 870\,$\mu$m size of the SMGs with this exponential kernel and find that the FWHM of the convolved profile is $3.8$--$5.2$\,kpc, showing that the radio sizes of \citet{Biggs08} are likely consistent with our 870\,$\mu$m sizes.

The galaxies presented by~\citet{Murphy08} are ``normal'' star-forming galaxies, with star formation rates three orders of magnitude lower than the SMGs presented here. The $e$-folding length of the kernel has been shown to depend on both star formation rate density and morphology, with high star formation rate densities and irregular galaxies having lower $e$-folding lengths~\citep{Murphy08}. In both of these situations ordered magnetic fields are disrupted, allowing cosmic ray electrons to stream more freely out of the galaxy, resulting in a shorter scale length. As such, we caution that a scale length of $1$--2\,kpc might be an over-estimate of the appropriate scale length for the radio emission from SMGs.

Next we compare the 870\,$\mu$m sizes of SMGs to the sizes of SMGs measured from $^{12}$CO emission lines. \citet{Engel10} present a compilation (including new observations) of the sizes of SMGs measured from resolved $^{12}$\,CO ($J$\,=\,3--2\,/\,4--3\,/\,6--5\,/\,7--6) emission lines (see also \citealt{Tacconi06,Tacconi08}). Deriving a median size for the sample is challenging, as the upper limits on the size of unresolved SMGs are typically similar to the sizes of the resolved sources. Treating the unresolved sources as point sources, the median size of the sample is FWHM\,$=$\,2.3\,$\pm$\,1.1\,kpc, but this rises to $3.7 \pm 0.6$\,kpc if upper limits are taken as the ``true'' source size. In our analysis we have not included the sizes taken from \citet{Bothwell10}, as it is not clear that these sizes have been deconvolved for the beam.

In the study by \citet{Engel10} the $^{12}$CO-emission lines are often either unresolved, or marginally resolved, with sizes constrained at only 1--2\,$\sigma$. The sizes are also measured from $^{12}$CO emission lines across a range of rotational transitions, from  $J$\,=\,2--1 to 7--6, and we note that there is ongoing debate over whether emission from different transitions traces the same molecular gas (\citealt{Weiss07,Panuzzo10,Bothwell13,Narayanan14}). Due to these issues we simply note that the sizes derived from {\it moderate}--$J$ transitions of $^{12}$CO are broadly consistent with the median $870$\,$\mu$m size of the SMGs in our sample.

Finally, we compare our results to the sizes of SMGs derived from high-resolution, {\it HST}, near--infrared imaging. \citet{Chen14} recently presented $H$--band/{\it HST} imaging (rest--frame optical imaging) of 48 SMGs. These SMGs were drawn from the ALESS sample \citep{Hodge13}, and have precise $\lsim 0.3''$ identifications from ALMA $870$\,$\mu$m observations. \citet{Chen14} fit a S\'{e}rsic profile to the $H$--band emission, deriving a median half--light radius of $R_{\rm e} =$\,4.4$^{+1.1}_{-0.5}$\,kpc (see also~\citealt{Targett13, Wiklind14}). As recommended by \citet{Chen14}, we have restricted their sample to the 17 SMGs with an apparent $H$--band magnitude $<$\,$24$\,mag and a photometric redshift between $z_{\rm   phot} =$\,1--3, which ensures that the size distribution is complete (median $S_{850}$\,=\,4.0\,$\pm$\,0.6\,mJy). The S\'{e}rsic profiles fitted by \citet{Chen14} are more complicated than the Gaussian profiles used in our analysis, but we note that a Gaussian profile corresponds to a S\'{e}rsic profile with S\'{e}rsic index $n=0.5$. If we convert our 870\,$\mu$m sizes to a half--light radius we find that the SMGs have a median $R_{\rm e} = 1.2 \pm 0.1$\,kpc in the sub--millimeter. As we show in \S\,\ref{subsubsec:sersic} the half--light radius of the best-fit S\'{e}rsic profile to the stacked $870$\,$\mu$m emission from the SMGs is consistent with the median size from the elliptical Gaussian fit. The optical sizes of the SMGs are about four times larger than the 870\,$\mu$m dust emitting region, and as can be seen in Figure~\ref{fig:hist} the distribution of near--infrared sizes has considerably more dispersion than the 870\,$\mu$m sizes. The compact nature of the star formation means that it is plausible we are observing bulge growth in the form of obscured star formation. However, to test this hypothesis requires high-resolution dust and optical imaging of the same sample of SMGs to pinpoint the location of the star formation within the host galaxy.

%
% Figure6- SFRd vs z
%
\begin{figure}
   \centerline{ \psfig{figure=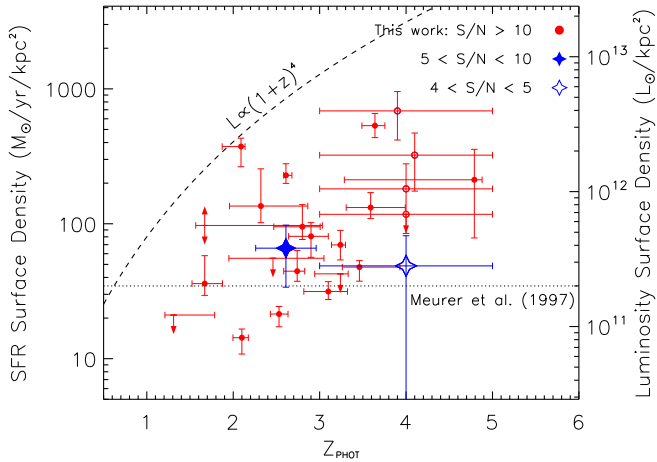,angle=90,width=0.48\textwidth}}
\caption{ Star formation rate surface density of the SMGs in our sample as a function of redshift. The dashed line represents luminosity evolution $L_{\rm IR} \propto (1+z)^4$. A dotted line shows the 90$^{th}$ percentile luminosity surface density for a sample of ultraviolet--selected sources \citep{Meurer97}. The SMGs have a median star formation rate density of  $90 \pm 30$\,$\Msol$\,yr$^{-1}$\,kpc$^{-2}$, and we note that no SMGs exceed the Eddington limit for a radiation pressure supported starburst ($\sim$\,1000\,$\Msol$\,yr$^{-1}$\,kpc$^{-2}$; \citealt{Andrews11})
}
 \label{fig:sfrd}
\end{figure}

\subsubsection{Comparison to local U/LIRGs}
We now compare the sizes of the SMGs in our sample to the sizes of local infrared--bright galaxies. Recently~\citet{Ueda14} presented a compilation of $<1$\,kpc resolution interferometric imaging (new and archival) of $^{12}$CO\,($J$\,=\,1--0; 2--1; 3--2) emission from 30 local star-forming galaxies classified as ``merger remnants''. We select galaxies from their sample with far-infrared luminosities $>$\,$1$\,$\times 10^{10}$\,$\Lsol$ (SFR\,$\gsim$\,1\,$\Msol$\,yr$^{-1}$), resulting in a sample of 24 galaxies with a median $L_{\rm {FIR}} = (1.4 \pm 0.1) \times 10^{11}$\,$\Lsol$. The sample contains one ULIRG ($L_{\rm {FIR}} > 10^{12}$\,$\Lsol$) and 12 LIRGs ($L_{\rm {FIR}} > 10^{11}$\,$\Lsol$). \citet{Ueda14} measure the size of each galaxy as the radius containing $80$\,\pc\, ($R_{80}$) of the total $^{12}$CO flux. To ensure a fair comparison with the results presented here, we derive the correction between FWHM and $R_{80}$, and apply this correction to the sizes presented by~\citet{Ueda14}. The sample of 24 local galaxies has a median diameter of $1.9 \pm 0.2$\,kpc, with a $1$\,$\sigma$ dispersion of $1.3$--3.4\,kpc (consistent with \citealt{Downes98}). 

We caution that \citet{Ueda14} present a sample of optically selected ``merger--remnants'' and although these ``merger--remnants'' are FIR--bright, the selection is not well--matched to the FIR--selected sample of SMGs presented here. \citet{Ueda14} also measure the size of the sources in their sample from {\it low}--$J$ transitions of $^{12}$CO: 12 sources are observed in $^{12}$CO\,($J$\,=\,1--0) emission, ten in $^{12}$CO\,($J$\,=\,2--1), and two in $^{12}$CO\,($J$\,=\,3--2). The lowest $J$ transitions trace cold molecular gas, and are expected to reflect the size of the total gas reservoir, rather than the star-forming region. Indeed, studies of Arp\,220, the closest ULIRG, find that the size of the rest--frame 860\,$\mu$m emission is 50--80\,pc, which is approximately two to four times smaller than the extent of the $^{12} $CO\,($J$\,=\,1--0) emission~\citep{Downes98, Sakamoto08}. 

The sizes of local U/LIRGs measured from {\it low}--$J$ transitions of $^{12}$CO appear marginally smaller than the 870\,$\mu$m sizes of the SMGs presented here, although the two samples are consistent within the associated uncertainties. However, as we have discussed, sizes measured from the $J$\,=\,1--0 transition of $^{12}$CO are likely to be an upper limit on size of the star-forming region. In the absence of a sample of local ULIRGs with sizes measured at approximately the same wavelength as the SMGs in this paper, we simply note that the star-forming regions in SMGs may be significantly larger than local U/LIRGs.

\subsection{Are SMGs Eddington Limited Starbursts?}
\label{sec:sfrd}  
The high star formation rates of SMGs means that they could host regions with an extreme star formation density, which could result in very different star formation conditions to other galaxy populations. In an isolated star-forming region the radiation pressure from massive stars may provide sufficient feedback to regulate the further collapse of the giant molecular cloud. \citet{Andrews11} show via the balance of radiation pressure from star formation with self-gravitation, that the maximum star formation rate surface density, assuming optically thick dust emission ($\tau$\,$>$\,1)\,\footnote{ \citet{Andrews11} state that a starburst should be treated as optically thick if the gas is above a critical surface density, 
  \begin{equation}
     \Sigma_{\rm g} \gsim 1200\Msol{\rm kpc}^{-2}\,.
 \end{equation}
 Here we have adopted a Rosseland mean dust opacity of $5$\,cm$^{-2}$\,g$^{-1}$ and a dust-to-gas ratio of $f_{\rm dg} = 90$ (see ~\citealt{Andrews11,Swinbank13}). Assuming the typical gas mass of SMGs, $4 \times 10^{10}$\,$\Msol$ \citep{Swinbank13}, then the typical gas density of SMGs is $\Sigma_{\rm g} = 4000$\,$\Msol$\,kpc$^{-2}$.}, and in the absence of nuclear heating via an active galactic nucleus (AGN), is
\begin{equation}
  {\rm SFR_{\rm max}} \gsim 11 f_{\rm gas}^{-0.5} f_{\rm dg}^{-1}\,\Msol
  {\rm yr}^{-1} {\rm kpc}^{-2}\,,
\end{equation}
where $f_{\rm gas}$ is the gas fraction in the star-forming region, and $f_{\rm dg}$ is the dust-to-gas ratio (see also~\citealt{Murray05,Thompson05,Murray10}). We adopt a dust--to--gas ratio of $f_{\rm dg} = 1/90$ \citep{Magnelli12,Swinbank13}, but note that $f_{\rm dg} = 1/50$ has also been suggested for star-forming galaxies~\citep{Kovacs06}. Estimates of the dynamical masses \citep{Swinbank06b,susie12}, stellar masses \citep{Hainline11,Simpson14} and gas masses \citep{Bothwell13} indicate that SMGs have typical gas fractions of $\sim 40$\,\pc; however, this is a global property of the galaxy. As the size of the rest--frame optical emitting region in SMGs is around four times larger than the size of the star--forming region (Figure~\ref{fig:hist}), it is likely that the gas fraction is considerably higher in the star-forming region. In our analysis we adopt a gas fraction of unity, i.e. gas is the dominant component, as this sets a lower limit on the maximum star formation rate surface density. If we adopt a gas fraction of $40$\,\pc\, then the maximum star formation rate surface density would rise by $60$\,\pc. Adopting these values the maximum star formation rate surface density predicted by~\citet{Andrews11} is $\sim 1000$\,$\Msol$\,yr$^{-1}$\,kpc$^{-2}$. 

The 23 SMGs in our sample have a median far-infrared luminosity of (5.7\,$\pm$\,0.7)\,$\times$10$^{12}$\,L$_\odot$, which corresponds to a median SFR\,=\,$990 \pm 120$\,$\Msol$\,yr$^{-1}$ (for a Salpeter IMF). To derive the star formation rate surface density for our sample we initially divide the SFR of each SMG by a factor of two (the measured size of each SMG corresponds to the half-light radius of the profile). We show our sample in Figure~6, and assuming a uniform disc profile for the gas and cool dust distribution the median star formation rate surface density of the SMGs is $90 \pm 30$\,$\Msol$\,yr$^{-1}$\,kpc$^{-2}$. Only two SMGs have a star formation rate surface density above $500$\,$\Msol$\,yr$^{-1}$\,kpc$^{-2}$, and no SMGs exceed the maximum value predicted by \citet{Andrews11}. 

The SMGs in our sample are forming stars at surface rate densities an order of magnitude lower than the estimated Eddington limit. However, we stress that these surface rate densities are integrated across the star-forming region (see \citealt{Meurer97}). In the well--studied lensed SMG SMM\,J2135-0102~\citep{Swinbank10Nature} it has been shown that the star formation is in four distinct ``clumps'' (FWHM\,=\,100--300\,pc), located within a gas disc with a half light radius $\sim$\,1\,kpc \citep{Swinbank11,Danielson11,Danielson13}. If the star formation in the SMGs in our sample is ``clumpy'', then individual regions in the gas disc may be Eddington limited, but the integrated star formation rate surface density will appear sub--Eddington. 

~\citet{Meurer97} show that a sample of ultraviolet--selected sources (at $z$\,=\,0--3) have a 90$^{\rm th}$ percentile global luminosity surface density of 2\,$\times$10$^{11}$\,L$_\odot$\,kpc$^{-2}$ (with an associated uncertainty of a factor of three), which they interpret as an upper limit on the global luminosity surface density of a starburst. The SMGs presented here have a median global luminosity surface density of (\,5.2\,$\pm$\,1.7)\,$\times$10$^{11}$\,L$_\odot$\,kpc$^{-2}$, a factor of three times higher than the value found by~\citet{Meurer97} (see Figure~6). A possible explanation for this discrepancy is that \citet{Meurer97} measure half--light radii for the galaxies in their sample from ultraviolet imaging, and assume that dust attenuation is uniform across the galaxy. As discussed in \S~\ref{subsubsec:multisizes} we find that the rest--frame optical half--light radii of SMG are a factor of four larger than the 870--$\mu$m dust half--light radius. The discrepancy in the optical sizes and dust sizes of SMGs indicates that their is significant differential reddening in these sources. If the ultraviolet--selected sources in \citep{Meurer97} suffer from similar differential reddening then the half--light radii measured in the ultra--violet are likely to overestimate the true extent of the star formation.

%
% Figure5- Histogram of sizes
%
\begin{figure}
   \centerline{\psfig{figure=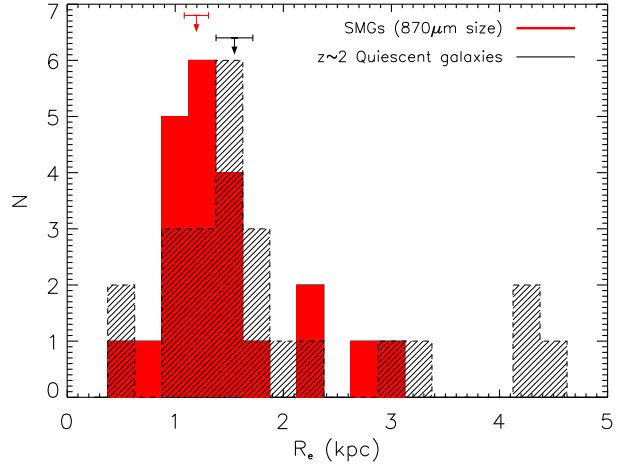,angle=90,width=0.48\textwidth}}
\caption{ Comparison of the FWHM of the rest--frame FIR emission in SMGs to $z \sim 2$ quiescent galaxies, the proposed progenitors of local elliptical galaxies. The quiescent galaxies are selected to have strong 4000\,\AA\,/\,Balmer breaks, suggestive of an old stellar population, and are found to be compact in F160W/{\it\,HST} imaging, with a median half light size of $R_{\rm e} = 1.5 \pm 0.2$, \citep{vandokkum08,Krogager13}. 
}

 \label{fig:hist_midz}
\end{figure}

\subsection{Are SMGs the progenitors of compact quiescent galaxies at $z \sim 2$?}
\label{sec:redndead}  
It has been suggested that an evolutionary sequence exists whereby SMGs transition into local elliptical galaxies via a quiescent galaxy phase at redshift $z$\,$\sim$\,2 (e.g.\ \citealt{Lilly99,Genzel03,Blain04a,Swinbank06b,Tacconi08,Swinbank10,Hickox12,Toft14,Simpson14,Chen14}). This evolutionary scenario has been investigated by comparing properties, such as the stellar masses, spatial clustering, and space densities of SMGs, to the properties of the proposed descendants. However, each of these methods has significant uncertainties: the stellar masses of SMGs have been shown to be highly dependent on the assumed star formation history; the number of SMGs with measured dynamical masses is small and the samples are inhomogeneous; and the spatial clustering strength of single-dish identified submm sources is biased by source blending. 

Here we use sizes to test this evolutionary link, since a distinctive feature of the population of quiescent galaxies at redshift $z$\,=\,1--3 is that they are extremely compact, with half-light sizes of $\sim$\,1\,kpc\,\footnote{We note that mass selected samples ($M_{\star} > 5$\,$\times$\,10$^{10}$\,M$_{\odot}$) containing both star--forming and quiescent galaxies at $z$\,=\,$1$--3,  have an half--light radius (4\,kpc) that is higher than, and a S\'{e}rsic index (1.5) which is lower than, our sample of quiescent galaxies (see~\citealt{Patel13,Buitrago13})} (e.g.\ \citealt{Daddi05,Zirm07,Toft07,Buitrago08, vandokkum08b,Newman12,Patel13,Krogager13}). A morphological analysis of SMGs thus provides an alternative route to investigating the proposed evolutionary sequence.

We construct a sample of spectroscopically confirmed quiescent galaxies at $z$\,$\sim$\,2, to compare to SMGs, by combining the samples presented by~\citet{vandokkum08} and \citet{Krogager13}. The galaxies in these samples have spectroscopically confirmed 4000\AA\,/\,Balmer breaks, indicative of an old ($\gsim$\,0.5--1\,Gyr) stellar population, assuming no age--dependent dust reddening. Both studies use the code {\sc Galfit} to fit a S{\'e}rsic profile to the $H$--band emission from each galaxy. The combined sample of 24 quiescent galaxies has a median half--light radius of $R_{\rm e}$\,=\,1.5\,$\pm$\,0.2\,kpc  and a median S{\'e}rsic index of $n$\,=\,3.3\,$\pm$\,0.7. The stellar masses of these galaxies are estimated to be $\gsim 10^{11}$\,$\Msol$ and the median redshift of the sample is $z$\,=\,2.2\,$\pm$\,0.1. The age of the stellar population in these quiescent galaxies (0.5--1\,Gyr) means that their progenitors likely lie at $z$\,=\,2.5--3  consistent with the redshift distribution of SMGs (\citealt{Toft14,Simpson14}). We note that the half--light radii of the spectroscopically confirmed quiescent galaxies in our comparison sample are consistent with the sizes measured for a large sample of color--selected quiescent galaxies at a similar redshift ($z$\,=\,1.5--2.5)~\citep{Patel13}.

As discussed in \S\,\ref{subsubsec:multisizes}, \citet{Chen14} show that the unobscured rest--frame optical light in SMGs has a median half light size of $R_{\rm e} =$\,4.4$^{+1.1}_{-0.5}$\,kpc and S\'{e}rsic index of $n$\,=\,1.2\,$\pm$\,$0.3$. In contrast, the quiescent galaxies in our comparison sample have a median half--light radius of $R_{\rm e}$\,=\,1.5\,$\pm$\,0.2\,kpc, significantly smaller than the rest--frame optical half--light radius of SMGs.  The median S{\'e}rsic index of the SMGs is also significantly lower than the quiescent galaxies, indicating that a significant transformation of the light profile (and potentially stellar mass distribution) of SMGs is required if they are indeed the progenitors of the quiescent galaxies in our comparison sample. A number of studies have suggested  that the star formation in SMGs is triggered by merger activity: in $H$--band/{\it HST} imaging (82\,$\pm$\,9)\,$\pc$\ of SMGs have signatures of interactions and/or irregular morphologies \citep{Chen14}, and the low S{\'e}rsic index $n$ measured for the population is consistent with merging systems \citep{Kartaltepe12}; dynamical studies of resolved H$\alpha$ emission from SMGs find strong evidence of kinematically distinct multiple components \citep{Swinbank06b,susie12}; and a study of molecular line emission from SMGs finds that 75\,\pc\ of sources are ongoing mergers, with the remaining 25\,\pc\ classed as either disc or late stage mergers~\citep{Engel10}. 

It is well established that mergers can induce torques that are effective at removing angular momentum from the gas in the system, allowing the gas to fall into the inner regions and potentially resulting in a kpc--scale nuclear starburst (e.g.\ \citealt{Barnes91,Mihos94}). Recently, \citet{Hopkins13} presented numerical simulations of two gas rich disc galaxies undergoing a major merger at high-redshift. \citet{Hopkins13} show that the major merger triggers an intense central starburst, with 50$\pc$\, of the star formation activity concentrated within a diameter of $\sim$\,$2$\,kpc. The simulations also demonstrate that the starburst increases the stellar surface density within the central 2\,kpc diameter by two orders of magnitude. The size of the star-forming region in the SMGs presented in this work are consistent with simulations of major-mergers. If the star formation in SMGs is indeed a centrally concentrated starburst then we expect a significant rise in the concentration of the stellar component, increasing the fraction of light contained in a de Vaucouleurs--like ($n$\,=\,4) profile.

\subsubsection{S\'{e}rsic index}
\label{subsubsec:sersic}
We now investigate whether there is evidence that the star formation in SMGs is more highly concentrated than the stellar component. First, we test whether our data have sufficiently high SNR to recover a more complex S\'{e}rsic profile\,\footnote{The test we perform is to investigate the required SNR, and does not test the effects of spatial filtering on our ALMA maps, due to the configuration of the array. However, we note that the maximum angular scale that our maps are sensitive to is 5$''$, which is an order of magnitude larger than the FWHM of the SMGs in our sample.}. We create 4000 model sources with a half-light size fixed at the median size of the SMGs, a range of S\'{e}rsic index from $n$\,=\,0.5--4 and an SNR distribution that is uniform from 10--30\,$\sigma$. To simulate realistic noise we add these models to a randomly chosen position in one of the residual (source-subtracted) ALMA maps. 

We fit a S\'{e}rsic profile to each model source using the code {\sc galfit} \citep{Peng10}, and find that the half-light size of the model sources is recovered accurately at all values of SNR and S\'{e}rsic index ($R_{\rm e}^{\rm rec}$\, /\,$R_{\rm e}^{\rm input} = $\,0.99\,$\pm$\,0.01; 1-$\sigma$ dispersion $\pm$\,0.2). In contrast, although the S\'{e}rsic index is recovered correctly on average ($n^{\rm rec}$\,/\,$n^{\rm input} = $\,0.97\,$\pm$\,0.02) the distribution has a 1-$\sigma$ dispersion of 0.4--2.2, indicating that measurements for individual sources are highly uncertain\,\footnote{If we re--fit our model sources, but force the S\'{e}rsic index to $n$\,=\,0.5 the median recovered half--light size is $R_{\rm e}^{\rm rec}$\, /\,$R_{\rm e}^{\rm input} = $\,0.99\,$\pm$\,0.01, with a 1-$\sigma$ dispersion $\pm$\,0.15}. Crucially, we find that the dispersion in the recovered S\'{e}rsic index is dependent on the SNR of the injected source; for SNR\,$>20$\,$\sigma$ the 1--$\sigma$ dispersion is 0.6--1.8. As such we do not attempt to fit a S{\'e}rsic profile on a source-by-source basis, but instead stack the 870\,$\mu$m emission from the 23 SMGs detected at $>$\,10\,$\sigma$ and use {\sc galfit} to fit a S\'{e}rsic model to the stacked profile. The best-fit model to the stacked profile has a S\'{e}rsic index of $n$\,=\,2.5\,$\pm$\,0.4 and a half-light radius of $R_{\rm e}$\,=\,0.17\,$\pm$\,0.03$''$,  where the errors are derived by bootstrapping the maps used in the stacking. 

The median S\'{e}rsic index of the 870\,$\mu$m emission from the SMGs is higher than the S\'{e}rsic index of the unobscured rest--frame optical light in SMGs (\citealt{Chen14}, but see also~\citealt{Targett13}), indicating that the ongoing star formation is more centrally concentrated. As we show in Figure~\ref{fig:hist_midz}, the half--light radius of the star formation in SMGs, median $R_{\rm e}$\,=\,1.2\,$\pm$\,0.1\,kpc, is comparable to the half--light radius of the quiescent galaxies in our comparison sample, $R_{\rm e}$\,=\,1.5\,$\pm$\,0.2\,kpc. If we combine the average S{\'e}rsic profile of the star formation and the pre--existing stellar population, under the simple assumption that the ongoing star-formation in SMGs doubles the luminosity of the stellar component, then the half--light radius of the resulting galaxy is $\sim$\,2\,kpc. This estimated half--light radius of the post--starburst SMGs is still marginally larger than median half--light radius of $z \sim 2$ quiescent galaxies. However, as shown by \citet{Chen14} (82\,$\pm$\,9)\,$\pc$\ of SMGs have signs of either merger activity or disturbed morphologies, and it is likely that the size measured for the ongoing merger is larger than the size of the galaxy at post-coalescence. In addition, the contribution of the stars formed in the SMG phase to the luminosity of the post--starburst galaxy is highly dependent on the mass of stars formed and the relative ages of the stars formed in the starburst and the pre--existing stellar population. The half--light radius of the ongoing, extreme, star--formation in SMGs is comparable to the half--light radius of high redshift compact quiescent galaxies, indicating that the ongoing star formation may explain the transformation required in the proposed evolutionary scenario from an intense starburst phase (SMG) to a local elliptical galaxies, via a compact quiescent galaxy at $z \sim 2$.

\section{Conclusions}
In this paper we have presented high-resolution ($0.3''$) ALMA imaging of 52 SMGs in the UDS field. The main conclusions from our work are: 

\begin{itemize}
\item We fit an elliptical Gaussian model to the 870\,$\mu$m emission from 23 SMGs detected at SNR\,$>$\,10\,$\sigma$. The median diameter, deconvolved from the beam, of these 23 SMGs is FWHM\,=\,$0.30 \pm 0.04''$, and the distribution has a narrow interquartile range of 0.26--0.42$''$. Two SMGs in our sample (10\,\%) are best fit by a point source model, and have an upper limit on their size of $<0.18''$. We stack the 870\,$\mu$m emission from SMGs detected at SNR\,=\,4--5\,$\sigma$ and SNR\,=\,5--10\,$\sigma$, measuring an average size of 0.35$^{+0.17}_{-0.10}$$''$ and $0.30 \pm 0.07''$ for each subset, respectively, consistent with the brighter examples.

\item Using photometric redshifts we convert the angular diameter of each SMG to a physical scale. The median physical FWHM of the 23 SMGs detected at SNR\,$>$\,10\,$\sigma$ is $2.4 \pm 0.2$\,kpc, and again the distribution has a narrow interquartile range of 1.8--3.2\,kpc. We investigate the size of the rest--frame FIR--emission region in SMGs, but do not find a trend with either redshift or 870\,$\mu$m flux density.

\item We compare the 870\,$\mu$m sizes of the SMG presented in this work to the sizes of SMGs measured at 1.4\,GHz. The median FWHM of these SMGs is $5.1 \pm 1.1$\,kpc, which is about two times larger than the median $870$\,$\mu$m size presented here. We convolve the median 870\,$\mu$m profile of the SMGs with an exponentially declining kernel, with a scale length of 1--2\,kpc, appropriate for modelling the radio and far--infrared emission in local star-forming galaxies \citep{Murphy08}. The convolved profile has a FWHM\,=\,3.8--5.2\,kpc, showing that the radio sizes may be consistent with our sub--millimeter sizes. However, we caution that as stated by \citet{Murphy08} the scale length of the convolution kernel may be sensitive to the star formation rate and morphology of the galaxy.

\item We also compare the sizes we derive for the 870\,$\mu$m emission region in SMGs to the sizes of SMGs measured from observations of resolved ``moderate''--$J$ $^{12}$\,CO emission lines. Treating the upper limits in the CO studies as point sources, the median FWHM for the sample is $2.3 \pm 1.1$\,kpc (rising to $3.7 \pm 0.6$\,kpc if upper limits are taken as detections). Thus, the cool gas estimates of the sizes are consistent with the 870\,$\mu$m sizes measured here.

\item The pre--existing stellar population in SMGs has a half--light radius of 4.4$^{+1.1}_{-0.5}$\,kpc, as measured from rest--frame optical {\it HST} imaging \citep{Chen14}, which is about four times larger than the extent of the ongoing star formation. The high dust content of SMGs means that the starburst component is likely to be missed in rest--frame optical imaging, while their prodigious star formation rates mean they have the potential to transform their stellar mass distribution.  Hence, we expect the post--starburst galaxy to be compact, with a smaller half--light radius than the pre-existing stellar population. 

\item SMGs are slightly larger than local far--infrared bright galaxies (U/LIRGs), with low--$J$ molecular gas sizes for the latter, median FWHM\,=\,$1.9 \pm 0.2$\,kpc.  However, these gas-derived sizes are likely to be larger than the far--infrared extents of the local galaxies, meaning the SMGs are probably physically larger systems.

\item The 23 SMGs in our sample have a median star formation rate of SFR\,=\,$1170\pm160$\,$\Msol$\,yr$^{-1}$ (for a Salpeter IMF), and a median star formation density of $90 \pm 30$\,$\Msol$\,yr$^{-1}$\,kpc$^{-2}$. The Eddington limit for a radiation pressure regulated starburst is $\sim 1000$\,$\Msol$\,yr$^{-1}$\,kpc$^{-2}$ \citep{Andrews11}. Hence, the SMGs in our sample have integrated star formation rate densities that are on average an order of magnitude lower than the Eddington limit. We suggest that this is due to the star formation occurring in ``clumps'', which may be Eddington limited, but appear sub--Eddington when integrated across the whole star formation region.

\item The half-light radius of the ongoing star formation in SMGs is similar to the size of $z$\,=\,$2.2$\,$\pm$\,0.1 quiescent galaxies, median $R_{\rm e}$\,=\,1.5\,$\pm$\,0.2. The ongoing, compact, starburst in SMGs has the potential to at least double the pre-existing stellar mass, offering an explanation of the proposed transformation of SMGs into compact, quiescent galaxies at high redshift, and thus how they fit into the evolutionary scenario proposed for local elliptical galaxies.

\end{itemize}

\section*{Acknowledgments}
We thank Adam Avison and the Manchester ALMA ARC node for their assistance verifying the calibration and imaging of our ALMA data. JMS and ALRD acknowledge the support of STFC studentships (ST/J501013/1 and ST/F007299/1, respectively). AMS acknowledges financial support from an STFC Advanced Fellowship (ST/H005234/1). IRS acknowledges support from the ERC Advanced Investigator program DUSTYGAL 321334, an RS/Wolfson Merit Award and STFC (ST/I001573/1). JSD acknowledges the support of the European Research Council via the award of an Advanced Grant, and the contribution of the EC FP7 SPACE project ASTRODEEP (312725). JEG acknowledges support from the Royal Society. RJI acknowledges support from the European Research Council (ERC) in the form of Advanced Grant, COSMICISM.

This paper makes use of the following ALMA data: ADS/JAO.ALMA$\#$2012.1.00090.S. ALMA is a partnership of ESO (representing its member states), NSF (USA) and NINS (Japan), together with NRC (Canada) and NSC and ASIAA (Taiwan), in cooperation with the Republic of Chile. The Joint ALMA Observatory is operated by ESO, AUI/NRAO and NAOJ. This publication also makes use of data taken with the SCUBA-2 camera on the James Clerk Maxwell Telescope. The James Clerk Maxwell Telescope is operated by the Joint Astronomy Centre on behalf of the Science and Technology Facilities Council of the United Kingdom, the National Research Council of Canada, and (until 31 March 2013) the Netherlands Organisation for Scientific Research. Additional funds for the construction of SCUBA-2 were provided by the Canada Foundation for Innovation. This research has made use of data from HerMES project (http://hermes.sussex.ac.uk/). HerMES is a Herschel Key Programme utilising Guaranteed Time from the SPIRE instrument team, ESAC scientists and a mission scientist.

All data used in this analysis can be obtained from the ALMA, ESO, {\it Spitzer} and {\it Herschel} archives.

\bibliographystyle{mn2e} 
\bibliography{ref.bib}

\end{document}